\documentclass[onecolumn,notitlepage,nofootinbib]{revtex4-1}
\usepackage{amsmath,amssymb,bm,graphicx,hyperref}
\allowdisplaybreaks[1]

\renewcommand\d{\partial}
\newcommand\grad{\bm\nabla}
\newcommand\+{\dagger}
\newcommand\<{\langle}
\renewcommand\>{\rangle}
\newcommand\up{\uparrow}
\newcommand\down{\downarrow}
\newcommand\eps{\varepsilon}
\renewcommand\k{{\bm{k}}}
\newcommand\K{{\bm{K}}}
\newcommand\p{{\bm{p}}}
\newcommand\q{{\bm{q}}}
\newcommand\x{{\bm{x}}}
\newcommand\y{{\bm{y}}}
\newcommand\C{\mathcal{C}}
\newcommand\E{\mathcal{E}}
\newcommand\M{\mathcal{M}}
\renewcommand\O{\mathcal{O}}
\renewcommand\P{\mathcal{P}}
\renewcommand\S{\mathcal{S}}
\newcommand\T{\mathcal{T}}
\newcommand\eq{\mathrm{eq}}
\newcommand\free{\mathrm{free}}
\newcommand\inter{\mathrm{int}}
\DeclareMathOperator\erf{erf}
\DeclareMathOperator\sgn{sgn}
\DeclareMathOperator\tr{tr}
\DeclareMathOperator\Tr{Tr}
\let\Re\relax\DeclareMathOperator\Re{Re}
\let\Im\relax\DeclareMathOperator\Im{Im}

\begin{document}

\title{Viscosity spectral functions of resonating fermions in the quantum virial expansion}

\author{Yusuke Nishida}
\affiliation{Department of Physics, Tokyo Institute of Technology,
Ookayama, Meguro, Tokyo 152-8551, Japan}

\date{April 2019}

\begin{abstract}
We consider two-component fermions with a zero-range interaction both in two and three dimensions and study their spectral functions of bulk and shear viscosities for an arbitrary scattering length.
Here the Kubo formulas are systematically evaluated up to the second order in the quantum virial expansion applicable to the high-temperature regime.
In particular, our computation of the bulk viscosity spectral function is facilitated by expressing it with the contact--contact response function, which can be measured experimentally under the periodic modulation of the scattering length.
The obtained formulas are fully consistent with the known constraints on high-frequency tail and sum rule.
Although our static shear viscosity agrees with that derived from the kinetic theory, our static bulk viscosity disagrees.
Furthermore, the latter for three dimensions exhibits an unexpected non-analyticity of $\zeta\sim(\ln a^2)/a^2$ in the unitarity limit $a\to\infty$, which thus challenges the ``crossover'' hypothesis.
\end{abstract}

\maketitle\twocolumngrid\tableofcontents\onecolumngrid

\section{Introduction, summary, and conclusion}
Transport coefficients provide valuable insights into strongly correlated systems complementary to thermodynamic quantities.
In particular, the shear viscosity has attracted significant interest across diverse subfields in physics~\cite{Schafer:2009,Adams:2012}, motivated by the universal lower bound on the shear viscosity to entropy density ratio~\cite{Kovtun:2005}.
Although the proposed conjecture was eventually violated~\cite{Brigante:2008a,Brigante:2008b,Kats:2009,Buchel:2009}, the quest for perfect fluids in strongly correlated quantum systems keeps on going.

Ultracold atoms offer ideal grounds to investigate strongly correlated quantum fluids because atomic interactions are tunable via Feshbach resonances~\cite{Chin:2010}.
Here the shear viscosity was measured for a strongly correlated Fermi gas of two components whose interaction saturates the unitarity bound~\cite{Cao:2011a,Cao:2011b}.
This measurement was later extended to the BCS-BEC crossover and the shear viscosity was found to be lower on the BEC side~\cite{Elliott:2014b}.
On the other hand, the bulk viscosity in the unitarity limit vanishes identically because of the conformal invariance~\cite{Son:2007}, which was also confirmed experimentally~\cite{Elliott:2014a}.
Furthermore, the bulk and shear viscosities were measured in the BCS-BEC crossover for a two-dimensional Fermi gas~\cite{Vogt:2012}.

The purpose of this paper is to present comprehensive analyses of viscosity spectral functions with full frequency dependence throughout the BCS-BEC crossover both in two and three dimensions.
We consider two-component fermions with a zero-range interaction parametrized only by the scattering length and evaluate the Kubo formulas systematically up to the second order in the quantum virial expansion, which is applicable to the high-temperature regime where the fugacity diminishes as $z\sim n/T^{d/2}$ at a fixed density.
To this end, the quantum virial expansion is reviewed in Sec.~\ref{sec:thermodynamics} for thermodynamic quantities, which is then applied to the spectral functions of bulk and shear viscosities in Secs.~\ref{sec:bulk} and \ref{sec:shear}, respectively.
In particular, our computation of the bulk viscosity spectral function is facilitated by expressing it with the contact--contact response function, which allows us to derive its formula even at $O(z^3)$ as shown in Appendix~\ref{appendix}.
We set $\hbar=k_B=1$ throughout this paper.

Because our analyses below appear to be involved, we first summarize our results and conclusions to guide readers who may not be interested in the intermediate technical details.
The spectral functions of bulk and shear viscosities up to the second order in fugacity are presented in Eqs.~(\ref{eq:bulk}) and (\ref{eq:shear}), respectively, which are valid for an arbitrary scattering length both in two and three dimensions.
We confirm in Eqs.~(\ref{eq:bulk_tail}), (\ref{eq:bulk_sum-rule}) for bulk and in Eqs.~(\ref{eq:shear_tail}), (\ref{eq:shear_sum-rule}) for shear that the obtained formulas are fully consistent with the known constraints on high-frequency tail and sum rule~\cite{Taylor:2010,Enss:2011,Hofmann:2011,Goldberger:2012,Taylor:2012}, which involves the contact density presented in Eq.~(\ref{eq:contact-density}).

The static limit of the shear viscosity spectral function requires the resummation, which leads to the static shear viscosity presented in Eq.~(\ref{eq:shear_static}) and agrees with that derived from the kinetic theory~\cite{Massignan:2005,Bruun:2005,Bruun:2012,Schafer:2012}.
On the other hand, the static limit of the bulk viscosity spectral function is presented in Eq.~(\ref{eq:bulk_static}) and turns out to disagree with that derived from the kinetic theory~\cite{Dusling:2013,Chafin:2013}.
Furthermore, our static bulk viscosity for three dimensions exhibits an unexpected non-analyticity of $\zeta\sim(\ln a^2)/a^2$ in the unitarity limit $a\to\infty$, which thus challenges the ``crossover'' hypothesis~\cite{Zwerger:2012}.
We also find that the static shear viscosity for three dimensions exhibits a weaker non-analyticity starting with $\eta\sim(\ln a^2)/a^6$ in the same limit.

The origin of these discrepancies between the Kubo formalism and the kinetic theory for the bulk viscosity is currently unknown and needs to be elucidated in a future study.
To this end, it may be useful to investigate higher order corrections to the bulk viscosity spectral function, which at the third order in fugacity is presented in Eq.~(\ref{eq:bulk_three-body}).
Also, as shown in Eqs.~(\ref{eq:bulk_contact-contact}), (\ref{eq:response_contact}), and (\ref{eq:response_energy-entropy}), the bulk viscosity spectral function can be extracted experimentally by measuring the contact, energy, or entropy density under the periodic modulation of the scattering length.
Therefore, ultracold atom experiments may be qualified to discriminate the two contradicting predictions from the Kubo formalism and the kinetic theory.

\section{Preliminaries}\label{sec:thermodynamics}
\subsection{Formulation}
In this paper, we consider two-component fermions with a zero-range interaction in $d$ spatial dimensions described by
\begin{align}\label{eq:hamiltonian}
\hat{H} = -\sum_{\sigma=\up,\down}\int\!d\x\,
\hat\psi_\sigma^\+(\x)\frac{\grad^2}{2m}\hat\psi_\sigma(\x)
+ \sum_{\sigma,\tau}\int\!d\x\,\hat\psi_\sigma^\+(\x)\hat\psi_\tau^\+(\x)\,
\frac{g}{2}\,\hat\psi_\tau(\x)\hat\psi_\sigma(\x),
\end{align}
where the annihilation and creation operators obey the usual anti-commutation relation of $\{\hat\psi_\sigma(\x),\hat\psi_\tau^\+(\y)\}=\delta_{\sigma\tau}\delta(\x-\y)$.
Before we work on the quantum virial expansion for viscosity spectral functions, let us warm up by reviewing that for the partition function.

The quantum virial expansion is a systematic expansion in terms of fugacity, which utilizes a simple fact that the grand canonical trace can be written as a sum of canonical traces~\cite{Liu:2013}.
In particular, the partition function reads
\begin{align}
Z = \Tr[e^{-\beta(\hat{H}-\mu\hat{N})}]
= \sum_{N=0}^\infty z^N\underbrace{\tr_N[e^{-\beta\hat{H}}]}_{Z_N},
\end{align}
where $z\equiv e^{\beta\mu}$ is the fugacity and obviously $Z_0=1$.
It is convenient to choose the complete set of $N$-body states to be momentum eigenstates,
\begin{align}
|\sigma_1\p_1,\dots,\sigma_N\p_N\>
= \frac1{\sqrt{N!}}\prod_{n=1}^N\hat\psi_{\sigma_n\p_n}^\+|0\>,
\end{align}
where $\hat\psi_{\sigma\p}^\+\equiv L^{-d/2}\int\!d\x\,e^{i\p\cdot\x}\,\hat\psi_\sigma^\+(\x)$ is the Fourier transform of the creation operator in a periodic box of volume $L^d$.
The canonical trace in the $N$-body sector is then provided by
\begin{align}
\tr_N[\hat\O] = \sum_{\{\sigma\}}\sum_{\{\p\}}
\<\sigma_1\p_1,\dots,\sigma_N\p_N|\hat\O|\sigma_1\p_1,\dots,\sigma_N\p_N\>.
\end{align}
Also, with the use of the identity operator,
\begin{align}
\openone_N = \sum_{\{\sigma\}}\sum_{\{\p\}}
|\sigma_1\p_1,\dots,\sigma_N\p_N\>\<\sigma_1\p_1,\dots,\sigma_N\p_N|,
\end{align}
the Hamiltonian~(\ref{eq:hamiltonian}) can be decomposed into distinct particle number sectors according to
\begin{align}
\hat{H} = \sum_{N=0}^\infty\underbrace{\openone_N\hat{H}\openone_N}_{\hat{H}_N}.
\end{align}

The one-body sector is trivial because the Hamiltonian is
\begin{align}
\hat{H}_1 = \sum_\sigma\sum_\p|\sigma\p\>\frac{\p^2}{2m}\<\sigma\p|
\end{align}
and does not involve the interaction.
Consequently, the partition function at $O(z)$ is simply provided by
\begin{align}
Z_1 = \tr_1[e^{-\beta\hat{H}_1}] = 2\sum_\p e^{-\beta\frac{\p^2}{2m}}.
\end{align}
On the other hand, the Hamiltonian in the two-body sector is
\begin{align}
\hat{H}_2 = \underbrace{\sum_{\sigma,\tau}\sum_{\p,\q}
|\sigma\p,\tau\q\>\frac{\p^2+\q^2}{2m}\<\sigma\p,\tau\q|}_{\hat{T}_2}
+ \underbrace{\sum_{\sigma,\tau}\sum_{\p,\q}\sum_{\p',\q'}|\sigma\p,\tau\q\>
\delta_{\p+\q,\p'+\q'}\frac{g}{L^d}\<\sigma\p',\tau\q'|}_{\hat{V}_2}.
\end{align}
In order to evaluate the corresponding Boltzmann operator, we write the Hamiltonian as a sum of kinetic and potential energy operators and utilize the identity of
\begin{align}
e^{-\beta\hat{H}_2} &= \int_{-\infty}^\infty\!\frac{dE}{\pi}\,e^{-\beta E}
\Im\!\left[\frac1{E-\hat{H}_2-i0^+}\right] \\
&= \int_{-\infty}^\infty\!\frac{dE}{\pi}\,e^{-\beta E}
\Im\!\left[\frac1{E-\hat{T}_2-i0^+} + \sum_{n=0}^\infty\frac1{E-\hat{T}_2-i0^+}
\hat{V}_2\left(\frac1{E-\hat{T}_2-i0^+}\hat{V}_2\right)^n\frac1{E-\hat{T}_2-i0^+}\right].
\end{align}
Because the summation over $n$ can be performed according to
\begin{align}
\sum_{n=0}^\infty\hat{V}_2\left(\frac1{E-\hat{T}_2-i0^+}\hat{V}_2\right)^n
&= \sum_{n=0}^\infty\sum_{\sigma,\tau}\sum_{\p,\q}\sum_{\p',\q'}
|\sigma\p,\tau\q\>\delta_{\p+\q,\p'+\q'}\frac{g}{L^d}
\left(\sum_\k\frac1{E-\frac{(\p+\q)^2}{4m}-\frac{\k^2}{m}-i0^+}\frac{g}{L^d}\right)^n
\<\sigma\p',\tau\q'| \\
&= \frac1{L^d}\sum_{\sigma,\tau}\sum_{\p,\q}\sum_{\p',\q'}
|\sigma\p,\tau\q\>\delta_{\p+\q,\p'+\q'}
\frac1{\frac1g-\frac1{L^d}\sum_\k\frac1{E-\frac{(\p+\q)^2}{4m}-\frac{\k^2}{m}-i0^+}}
\<\sigma\p',\tau\q'|,
\end{align}
the Boltzmann operator is found to be
\begin{subequations}\label{eq:boltzmann}
\begin{align}
e^{-\beta\hat{H}_2} = e^{-\beta\hat{H}_2}\big|_\free + e^{-\beta\hat{H}_2}\big|_\inter,
\end{align}
with its free and interacting parts provided by
\begin{align}
e^{-\beta\hat{H}_2}\big|_\free = \sum_{\sigma,\tau}\sum_{\p,\q}
|\sigma\p,\tau\q\>e^{-\beta\frac{\p^2+\q^2}{2m}}\<\sigma\p,\tau\q|
\end{align}
and
\begin{align}\label{eq:interaction}
e^{-\beta\hat{H}_2}\big|_\inter &= \frac1{L^d}\sum_{\sigma,\tau}\sum_{\p,\q}\sum_{\p',\q'}
|\sigma\p,\tau\q\>e^{-\beta\frac{(\p+\q)^2}{4m}}\delta_{\p+\q,\p'+\q'} \notag\\
&\quad \times \int_{-\infty}^\infty\!\frac{d\eps}{\pi}\,e^{-\beta\eps}
\Im\!\left[\frac{\T_2(\eps-i0^+)}{(\eps-\frac{(\p-\q)^2}{4m}-i0^+)(\eps-\frac{(\p'-\q')^2}{4m}-i0^+)}\right]
\<\sigma\p',\tau\q'|,
\end{align}
\end{subequations}
respectively.
Here the integration variable is changed to $\eps=E-(\p+\q)^2/4m$ and
\begin{align}
\T_2(\eps-i0^+) \equiv \frac1{\frac1g - \frac1{L^d}\sum_\k\frac1{\eps-\frac{\k^2}{m}-i0^+}}
\end{align}
is the two-body scattering $T$-matrix in the center-of-mass frame.
We note that Eq.~(\ref{eq:boltzmann}) is valid for any complex $\beta$ and the interacting part (\ref{eq:interaction}) vanishes when $\sigma=\tau$ for fermions.
Consequently, the partition function at $O(z^2)$ is provided by
\begin{align}
Z_2 = \tr_2[e^{-\beta\hat{H}_2}]
= \underbrace{2\sum_{\p,\q}e^{-\beta\frac{\p^2+\q^2}{2m}}}_{Z_1^2/2}
- \sum_{\p,\q}e^{-\beta\frac{\p^2+\q^2}{2m}}\delta_{\p\q}
+ \frac1{L^d}\sum_{\K,\k}e^{-\beta\frac{\K^2}{4m}}
\int_{-\infty}^\infty\!\frac{d\eps}{\pi}\,e^{-\beta\eps}
\Im\!\left[\frac{\T_2(\eps-i0^+)}{(\eps-\frac{\k^2}{m}-i0^+)^2}\right],
\end{align}
where $\K=\p+\q$ and $\k=(\p-\q)/2$ are the center-of-mass and relative momenta, respectively.

\subsection{Thermodynamics}
From the above quantum virial expansion for the partition function, we obtain that for the pressure according to
\begin{align}
\frac{p}{T} = \frac{\ln Z}{L^d}
&= \frac1{L^d}\left[zZ_1 + z^2Z_2 - z^2\frac{Z_1^2}{2} + O(z^3)\right] \\
&= \frac{2z}{L^d}\sum_\p e^{-\beta\frac{\p^2}{2m}}
- \frac{z^2}{L^d}\sum_\p e^{-\beta\frac{\p^2}{m}}
+ \frac{z^2}{L^{2d}}\sum_{\K,\k}e^{-\beta\frac{\K^2}{4m}}
\int_{-\infty}^\infty\!\frac{d\eps}{\pi}\,e^{-\beta\eps}
\Im\!\left[\frac{\T_2(\eps-i0^+)}{(\eps-\frac{\k^2}{m}-i0^+)^2}\right] + O(z^3).
\end{align}
Finally, by taking the thermodynamic limit, $L^{-d}\sum_\p\to\int\!d\p/(2\pi)^d$, and performing the resulting momentum integrations, we find
\begin{align}\label{eq:pressure}
\frac{p}{T} = \frac{2z}{\lambda_T^d} - \frac{z^2}{2^{d/2}\lambda_T^d}
+ \frac{2^{d/2}z^2}{2\Omega_{d-1}\lambda_T^d}
\int_{-\infty}^\infty\!\frac{d\eps}{\pi}\,e^{-\eps/T}
\Im\!\left[\frac{m^2\T_2(\eps-i0^+)}{(-m\eps+i0^+)^{2-d/2}}\right] + O(z^3),
\end{align}
where $\lambda_T\equiv\sqrt{2\pi/mT}$ is the thermal de Broglie wavelength and $\Omega_{d-1}\equiv(4\pi)^{d/2}/2\Gamma(2-d/2)=2,2\pi,4\pi$ coincides with the surface area of the unit $(d-1)$-sphere for $d=1,2,3$.
Similarly, the two-body scattering $T$-matrix in the same limit is found to be
\begin{align}
\T_2(\eps-i0^+) = \frac{\Omega_{d-1}}{m}\frac{d-2}{a^{2-d} - (-m\eps+i0^+)^{d/2-1}},
\end{align}
where $a$ is the scattering length introduced via
\begin{align}
\frac1g = \frac{ma^{2-d}}{(d-2)\Omega_{d-1}}
\end{align}
in the dimensional regularization~\cite{Fujii:2018}.

The partial derivative of the pressure with respect to the scattering length then leads to the so-called contact density~\cite{Tan:2008a,Tan:2008b,Tan:2008c},
\begin{align}\label{eq:contact-density}
\C = -\Omega_{d-1}ma^{d-1}\frac{\d p}{\d a}
= \frac{2^{d/2}z^2}{\lambda_T^d}
\int_{-\infty}^\infty\!\frac{d\eps}{\pi}\,e^{-\eps/T}\Im[m^2\T_2(\eps-i0^+)] + O(z^3),
\end{align}
where the integration by parts following
\begin{align}
a^{d-1}\frac{\d\T_2(\eps-i0^+)}{\d a}
= -\frac{2(-m\eps+i0^+)^{2-d/2}}{m}\frac{\d\T_2(\eps-i0^+)}{\d\eps}
\end{align}
is utilized.
Because the imaginary part of the two-body scattering $T$-matrix is provided by
\begin{align}\label{eq:T-matrix}
\Im[\T_2(\eps-i0^+)]
= \theta(a)\,\frac{2\Omega_{d-1}}{m^2a^{4-d}}\,\pi\delta(\eps+\tfrac1{ma^2})
+ \theta(\eps) \times
\begin{cases}\displaystyle
\frac{2\pi}{m}\frac{2\pi}{\ln^2(ma^2\eps)+\pi^2} &\quad (d=2),
\medskip\\\displaystyle
\frac{4\pi}{m}\frac{\sqrt{m\eps}}{m\eps+\frac1{a^2}} &\quad (d=3),
\end{cases}
\end{align}
the contact density for $d=3$ has the analytic expression of~\cite{Ho:2004,Yu:2009}
\begin{align}
\C = \frac{16\pi z^2}{\lambda_T^4}\left[1 + \sqrt\pi\,\frac{1+\erf(\tilde{a}^{-1})}{\tilde{a}}
\exp(\tilde{a}^{-2})\right]_{\tilde{a}=\sqrt{2\pi}a/\lambda_T} + O(z^3).
\end{align}
In particular, it is smooth even in the unitarity limit at $\lambda_T/a=0$ so as to be consistent with the ``crossover'' hypothesis~\cite{Zwerger:2012}.

\section{Bulk viscosity}\label{sec:bulk}
\subsection{Formulation}
We now turn to the quantum virial expansion for viscosity spectral functions.
In terms of the stress--stress response function,
\begin{align}
\chi_{ij,kl}(\omega) = \frac{i}{ZL^d}\int_0^\infty\!dt\,e^{i(\omega+i0^+)t}
\Tr[[\hat\Pi_{ij}(t),\hat\Pi_{kl}(0)]e^{-\beta(\hat{H}-\mu\hat{N})}],
\end{align}
the spectral functions of bulk and shear viscosities are defined according to~\cite{Taylor:2010}
\begin{align}\label{eq:viscosities}
\frac{\Im[\chi_{ij,kl}(\omega)]}{\omega}
= \left[\zeta(\omega)-\frac2d\,\eta(\omega)\right]\delta_{ij}\delta_{kl}
+ \eta(\omega)\left(\delta_{ik}\delta_{jl}+\delta_{il}\delta_{jk}\right).
\end{align}
Here $\hat\Pi_{ij}(t)=e^{i\hat{H}t}\hat\Pi_{ij}e^{-i\hat{H}t}$ is the stress tensor operator in the Heisenberg picture and its trace is directly related to the bulk viscosity spectral function via
\begin{align}
\zeta(\omega) = \frac1{d^2}\sum_{i,j=1}^d\frac{\Im[\chi_{ii,jj}(\omega)]}{\omega}.
\end{align}

For two-component fermions described by the Hamiltonian~(\ref{eq:hamiltonian}), the stress tensor operator assuming the dimensional regularization is provided by~\cite{Goldberger:2012,Fujii:2018}
\begin{align}\label{eq:stress}
\hat\Pi_{ij} = -\sum_{\sigma=\up,\down}\int\!d\x\,
\hat\psi_\sigma^\+(\x)\frac{\d_i\d_j}{m}\hat\psi_\sigma(\x)
+ \sum_{\sigma,\tau}\int\!d\x\,\hat\psi_\sigma^\+(\x)\hat\psi_\tau^\+(\x)\,
\frac{g}{2}\delta_{ij}\,\hat\psi_\tau(\x)\hat\psi_\sigma(\x).
\end{align}
Its trace is then simplified into
\begin{align}
\sum_{i=1}^d\hat\Pi_{ii} = 2\hat{H} + \frac{\hat{C}}{\Omega_{d-1}ma^{d-2}},
\end{align}
where the last term quantifies the conformal symmetry breaking with
\begin{align}\label{eq:contact}
\hat{C} = \sum_{\sigma,\tau}\int\!d\x\,\hat\psi_\sigma^\+(\x)\hat\psi_\tau^\+(\x)
\frac{(mg)^2}{2}\hat\psi_\tau(\x)\hat\psi_\sigma(\x)
\end{align}
being the contact operator~\cite{Braaten:2008,Barth:2011,Hofmann:2012}.
Because the commutator of the Hamiltonian with any operator in the grand canonical average vanishes, the bulk viscosity spectral function turns into the favorite form of
\begin{align}\label{eq:bulk_contact-contact}
\zeta(\omega) = \frac1{(d\,\Omega_{d-1}ma^{d-2})^2}\frac{\Im[\chi_{CC}(\omega)]}{\omega},
\end{align}
where
\begin{align}\label{eq:contact-contact}
\chi_{CC}(\omega) = \frac{i}{ZL^d}\int_0^\infty\!dt\,e^{i(\omega+i0^+)t}
\Tr[[\hat{C}(t),\hat{C}(0)]e^{-\beta(\hat{H}-\mu\hat{N})}]
\end{align}
is the contact--contact response function~\cite{Martinez:2017,Fujii:2018}.

The quantum virial expansion for the bulk viscosity spectral function is thus to evaluate
\begin{align}
\chi_{CC}(\omega)
= \sum_{N=0}^\infty\frac{z^N}{Z}\underbrace{\frac{i}{L^d}\int_0^\infty\!dt\,e^{i(\omega+i0^+)t}
\tr_N[e^{-\beta\hat{H}+i\hat{H}t}\hat{C}e^{-i\hat{H}t}\hat{C}
- e^{-\beta\hat{H}-i\hat{H}t}\hat{C}e^{i\hat{H}t}\hat{C}]}_{\chi_{CC}^{(N)}(\omega)}.
\end{align}
Here the contact operator~(\ref{eq:contact}) can be decomposed into distinct particle number sectors according to
\begin{align}
\hat{C} = \sum_{N=0}^\infty\underbrace{\openone_N\hat{C}\openone_N}_{\hat{C}_N},
\end{align}
where nontrivial matrix elements first appear from the two-body sector in the form of
\begin{align}
\hat{C}_2 = \sum_{\sigma,\tau}\sum_{\p,\q}\sum_{\p',\q'}|\sigma\p,\tau\q\>
\delta_{\p+\q,\p'+\q'}\frac{(mg)^2}{L^d}\<\sigma\p',\tau\q'|.
\end{align}
On the other hand, the time evolution operator such as $e^{-i\hat{H}_2t}$ is obtained from Eq.~(\ref{eq:boltzmann}) with the simple replacement of $\beta\to it$.
Because the bare coupling constant vanishes as $mg\to-2\pi/\ln\Lambda$ for $d=2$ and as $mg\to-2\pi^2/\Lambda$ for $d=3$ in the cutoff regularization with $|\k|<\Lambda$, the free part of the time evolution operator $e^{-i\hat{H}_2t}|_\free$ has vanishing contributions in the limit of $\Lambda\to\infty$ when sandwiched by the contact operators.
Therefore, only the interacting part of the time evolution operator $e^{-i\hat{H}_2t}|_\inter$ has non-vanishing contributions to the contact--contact response function.
By utilizing the identity of
\begin{align}\label{eq:identity}
\frac{g}{L^d}\sum_\k\frac1{\eps-\frac{\k^2}{m}-i0^+}
= 1 - \frac{g}{\T_2(\eps-i0^+)} \underset{\Lambda\to\infty}\to 1 \qquad (d\geq2),
\end{align}
we obtain
\begin{align}
\tr_2[e^{-\beta\hat{H}_2+i\hat{H}_2t}\hat{C}_2e^{-i\hat{H}_2t}\hat{C}_2]
= \sum_\K e^{-\beta\frac{\K^2}{4m}}
\int_{-\infty}^\infty\!\frac{d\eps}{\pi}\int_{-\infty}^\infty\!\frac{d\eps'}{\pi}\,
e^{-\beta\eps+i\eps t-i\eps't}\Im[m^2\T_2(\eps-i0^+)]\Im[m^2\T_2(\eps'-i0^+)].
\end{align}
Consequently, the contact--contact response function at $O(z^2)$ is found to be
\begin{align}\label{eq:bulk_two-body}
\chi_{CC}^{(2)}(\omega) = -\frac1{L^d}\sum_\K e^{-\beta\frac{\K^2}{4m}}
\int_{-\infty}^\infty\!\frac{d\eps}{\pi}\int_{-\infty}^\infty\!\frac{d\eps'}{\pi}
\frac{e^{-\beta\eps}-e^{-\beta\eps'}}{\eps-\eps'+\omega+i0^+}
\Im[m^2\T_2(\eps-i0^+)]\Im[m^2\T_2(\eps'-i0^+)],
\end{align}
from which the bulk viscosity spectral function in the thermodynamic limit reads
\begin{align}\label{eq:bulk}
\zeta(\omega) = \frac{2^{d/2}z^2}{(d\,\Omega_{d-1}a^{d-2})^2\lambda_T^d}
\frac{1-e^{-\omega/T}}{\omega}\int_{-\infty}^\infty\!\frac{d\eps}{\pi}\,
e^{-\eps/T}\Im[m\T_2(\eps-i0^+)]\Im[m\T_2(\eps+\omega-i0^+)] + O(z^3).
\end{align}
The higher order correction at $O(z^3)$ is presented in Eq.~(\ref{eq:bulk_three-body}).

\subsection{Evaluation}
Our concise formula for the bulk viscosity spectral function is even under $\omega\to-\omega$ and has the high-frequency tail of
\begin{align}\label{eq:bulk_tail}
\lim_{\omega\to\infty}\zeta(\omega)
= \frac{\C}{(d\,\Omega_{d-1}a^{d-2})^2} \times
\begin{cases}\displaystyle
\frac{4\pi^2}{|m\omega|\ln^2|ma^2\omega|} &\quad (d=2),
\medskip\\\displaystyle
\frac{4\pi}{|m\omega|^{3/2}} &\quad (d=3),
\end{cases}
\end{align}
where $\C$ is the contact density from Eq.~(\ref{eq:contact-density}).
This asymptotic form agrees with Refs.~\cite{Hofmann:2011,Goldberger:2012} for $d=3$ and with Ref.~\cite{Hofmann:2011} for $d=2$.%
\footnote{Although Eq.~(62) of Ref.~\cite{Hofmann:2011} presents a different result for $d=2$, the substitution of Eqs.~(45), (47) into Eqs.~(57), (58) in Ref.~\cite{Hofmann:2011} indeed leads to the same result as our Eq.~(\ref{eq:bulk_tail}).}
Also, the integration over $\omega$ leads to
\begin{align}
\int_{-\infty}^\infty\!\frac{d\omega}{\pi}\,\zeta(\omega)
= \frac{2^{1+d/2}z^2}{(d\,\Omega_{d-1}a^{d-2})^2\lambda_T^d}
\int_{-\infty}^\infty\!\frac{d\eps}{\pi}\int_{-\infty}^\infty\!\frac{d\omega}{\pi}
\frac\P{\omega-\eps}\,e^{-\eps/T}\Im[m\T_2(\eps-i0^+)]\Im[m\T_2(\omega-i0^+)] + O(z^3).
\end{align}
With the use of the Kramers--Kronig relation followed by the identity of
\begin{align}
2\Im[m\T_2(\eps-i0^+)]\Re[m\T_2(\eps-i0^+)] = \Omega_{d-1}a^{d-1}\frac\d{\d a}\Im[m\T_2(\eps-i0^+)],
\end{align}
we obtain
\begin{align}\label{eq:bulk_sum-rule}
\int_{-\infty}^\infty\!\frac{d\omega}{\pi}\,\zeta(\omega)
= -\frac{a^{3-d}}{d^2\Omega_{d-1}m}\frac{\d\C}{\d a},
\end{align}
which is consistent with the sum rule derived in Ref.~\cite{Taylor:2010} for $d=3$ and in Ref.~\cite{Taylor:2012} for $d=2$.
We note that the partial derivative with respect to the scattering length at fixed entropy and particle number is equivalent to that at fixed $T$ and $\mu$ to the lowest order in fugacity.

\begin{figure}[t]
\includegraphics[width=0.46\textwidth]{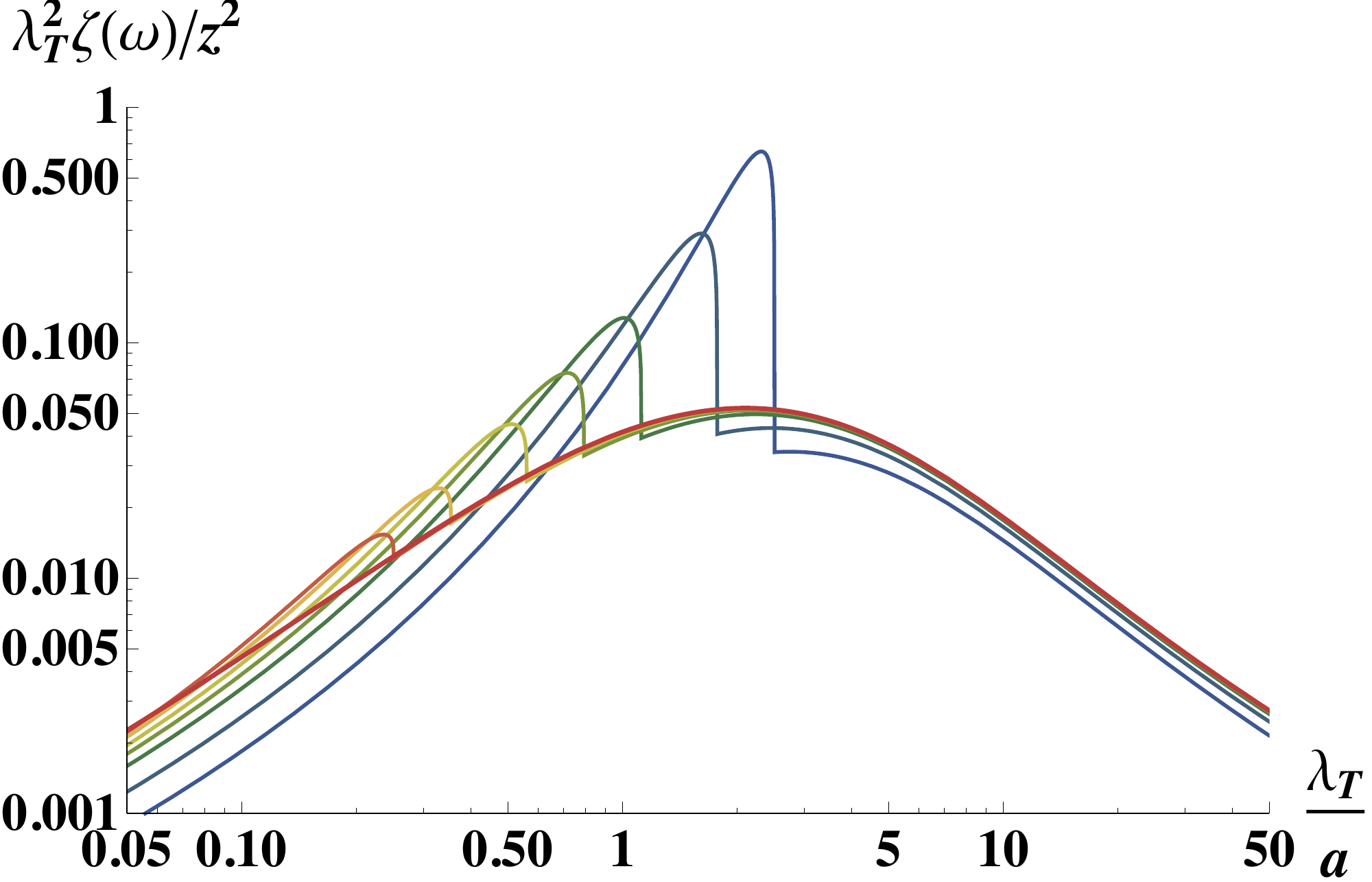}\hfill
\includegraphics[width=0.09\textwidth]{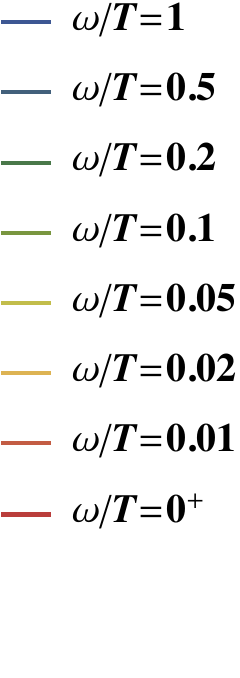}\hfill
\includegraphics[width=0.45\textwidth]{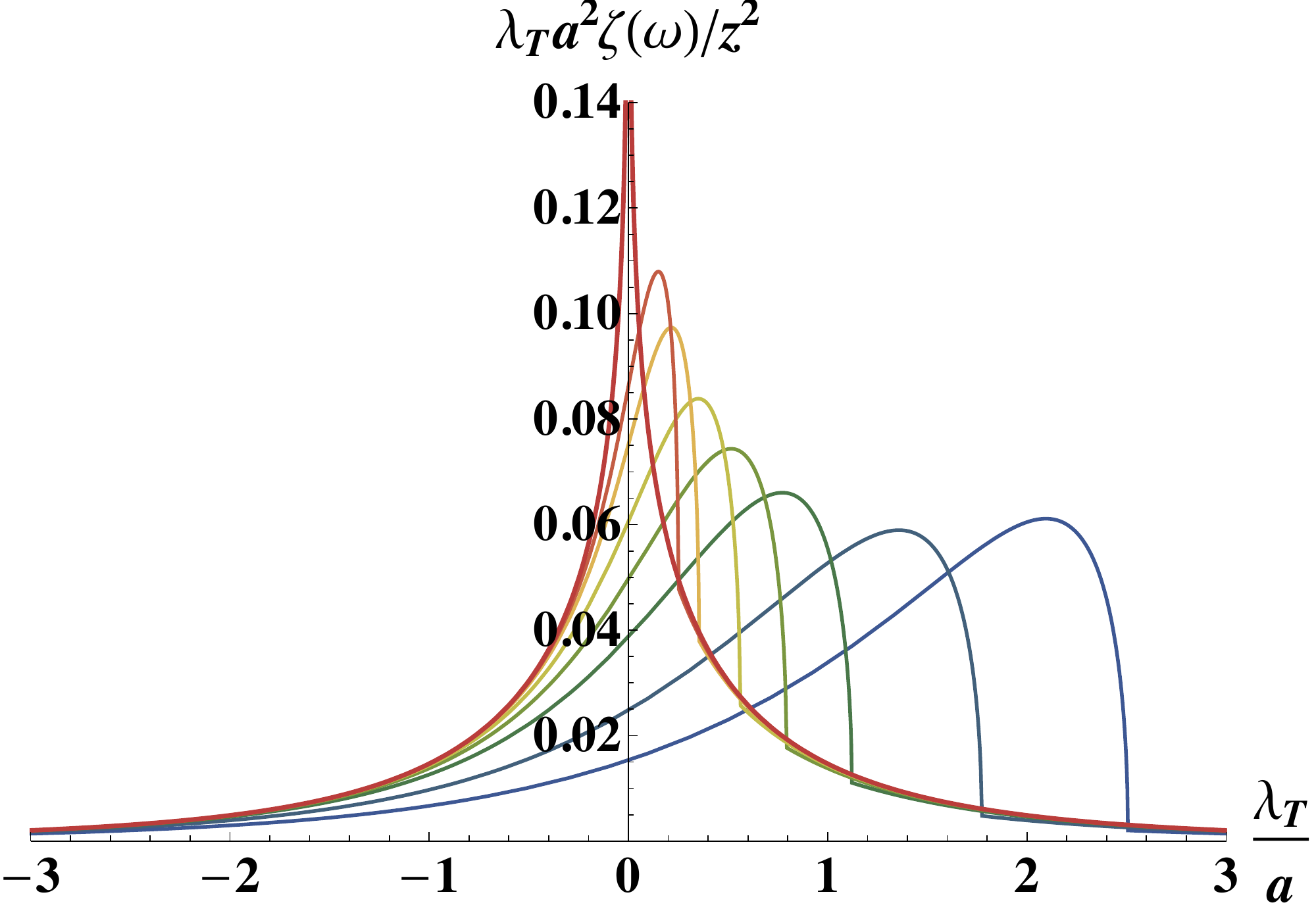}
\caption{\label{fig:bulk}
Static and dynamic bulk viscosities from Eq.~(\ref{eq:bulk_dynamic}) as functions of $\lambda_T/a$ in the forms of $\lambda_T^2\zeta(\omega)/z^2$ for $d=2$ (left panel) and of $\lambda_Ta^2\zeta(\omega)/z^2$ for $d=3$ (right panel) with the frequencies chosen at $\omega/T=0^+,0.01,0.02,0.05,0.1,0.2,0.5,1$.}
\end{figure}

After confirming our formula (\ref{eq:bulk}) in light of the known constraints, we are ready to evaluate it more closely.
The substitution of Eq.~(\ref{eq:T-matrix}) into Eq.~(\ref{eq:bulk}) leads to
\begin{align}\label{eq:bulk_dynamic}
\frac{d^2\lambda_T^d}{2^{d/2}z^2}\zeta(\omega)
&= \frac{1-e^{-|\omega|/T}}{|\omega|}e^{1/ma^2T}\frac2{ma^2}
\left[\frac2{ma^2}\pi\delta(\omega) + \rho_2(|\omega|-\tfrac1{ma^2})\,
\theta(|\omega|-\tfrac1{ma^2})\right]\theta(a) \notag\\
&\quad + \frac{1-e^{-|\omega|/T}}{|\omega|}\int_0^\infty\!\frac{d\eps}{\pi}\,
e^{-\eps/T}\rho_2(\eps)\rho_2(\eps+|\omega|) + O(z),
\end{align}
where the imaginary part of the two-body scattering $T$-matrix is normalized according to
\begin{align}
\rho_2(\eps) \equiv \left.\frac{\Im[m\T_2(\eps-i0^+)] }{\Omega_{d-1}a^{d-2}}\right|_{\eps>0}
=
\begin{cases}\displaystyle
\frac{2\pi}{\ln^2(ma^2\eps)+\pi^2} &\quad (d=2),
\medskip\\\displaystyle
\frac{\sgn(a)\sqrt{ma^2\eps}}{ma^2\eps+1} &\quad (d=3).
\end{cases}
\end{align}
The three terms in Eq.~(\ref{eq:bulk_dynamic}) correspond to bound--bound, bound--continuum, and continuum--continuum transitions, which contribute to the bulk viscosity spectral function at $\omega=0$, $|\omega|>1/ma^2$, and $\forall\omega$, respectively.
Therefore, the first two terms contribute only when the two-body bound state exists, $a>0$, which is always the case for $d=2$.
The resulting dynamic bulk viscosity is plotted in Fig.~\ref{fig:bulk} as well as its static limit, $\zeta\equiv\lim_{\omega\to0^+}\zeta(\omega)$, which excluding the singular term of $\delta(\omega)\theta(a)$ is provided by
\begin{align}\label{eq:bulk_static}
\lambda_T^d\zeta =
\begin{cases}\displaystyle
\frac{z^2}{2}\int_0^\infty\!\frac{d\tilde\eps}{\pi}\,e^{-\tilde\eps}
\left[\frac{2\pi}{\ln^2(\tilde{a}^2\tilde\eps)+\pi^2}\right]^2 + O(z^3) &\quad (d=2),
\medskip\\\displaystyle
\frac{2\sqrt2z^2}{9\pi}\frac{-1 + (1+\tilde{a}^{-2})
\exp(\tilde{a}^{-2})\Gamma(0,\tilde{a}^{-2})}{\tilde{a}^2} + O(z^3) &\quad (d=3).
\end{cases}
\end{align}
Here the integration variable is changed to $\tilde\eps=\eps/T$ and the dimensionless scattering length is introduced via $\tilde{a}\equiv\sqrt{mT}a=\sqrt{2\pi}a/\lambda_T$.

The static bulk viscosity for $d=2$ is a smooth function of $\lambda_T/a>0$ and has the asymptotic forms of
\begin{align}
\lim_{a\to0,\infty}\lambda_T^2\zeta
= \frac{2\pi z^2}{\ln^4(e^{-\gamma}\tilde{a}^2)} + O(z^2\ln^{-6}\tilde{a}^2,z^3).
\end{align}
On the other hand, the static bulk viscosity for $d=3$ has the asymptotic forms of
\begin{align}
\lim_{a\to0}\lambda_T^3\zeta
= \frac{2\sqrt2z^2}{9\pi}\tilde{a}^2 + O(z^2\tilde{a}^4,z^3)
\end{align}
and
\begin{align}
\lim_{a\to\infty}\lambda_T^3\zeta = \frac{2\sqrt2z^2}{9\pi}
\frac{\ln(e^{-1-\gamma}\tilde{a}^2)}{\tilde{a}^2} + O(z^2\tilde{a}^{-4},z^3),
\end{align}
which turns out to exhibit the non-analyticity in the unitarity limit as shown in the right panel of Fig.~\ref{fig:bulk}.
Our results disagree with the static bulk viscosity derived from the kinetic theory, where reported are $\lim_{a\to\infty}\lambda_T^3\zeta=z^2/(12\sqrt2\tilde{a}^2)$ for $d=3$~\cite{Dusling:2013} and $\lim_{a\to\infty}\lambda_T^2\zeta=z^2/(2\pi\ln^4\tilde{a}^2)$ for $d=2$~\cite{Chafin:2013}.
The origin of these discrepancies is currently unknown.
We also note that the dynamic bulk viscosity for $d=3$ toward the unitarity limit is provided by
\begin{align}
\lim_{a\to\infty}\lambda_T^3\zeta(\omega)
= \frac{2\sqrt2z^2}{9\pi\tilde{a}^2}\frac{2T}{\omega}
\sinh\!\left(\frac\omega{2T}\right)K_0\!\left(\frac\omega{2T}\right) + O(z^2\tilde{a}^{-3},z^3),
\end{align}
so that the limits of $\omega\to0$ and $a\to\infty$ do not commute mutually.

The bulk viscosity spectral function can be measured experimentally via the contact--contact response function~(\ref{eq:contact-contact}).
When the scattering length is periodically modulated as $a(t)=a+\delta a\sin(\omega t)$, the linear response theory predicts that the contact density responds according to~\cite{Fujii:2018}
\begin{align}\label{eq:response_contact}
\C(t) - \C_\eq[a(t)]
= \frac{\Im[\chi_{CC}(\omega)]}{\Omega_{d-1}ma^{d-1}}\,\delta a\cos(\omega t)
- \frac{\Re[\chi_{CC}(\omega)]-\chi_{CC}(0)}{\Omega_{d-1}ma^{d-1}}\,\delta a\sin(\omega t)
+ O(\delta a^2),
\end{align}
where $\C_\eq[a]\equiv\Tr[\hat{C}e^{-\beta(\hat{H}-\mu\hat{N})}]/(ZL^d)$ is the contact density in thermodynamic equilibrium for the scattering length $a$ and $\chi_{CC}(-\omega)=\chi_{CC}^*(\omega)$ is utilized.
Then, the thermodynamic identity together with the adiabatic and dynamic sweep theorems~\cite{Tan:2008a,Tan:2008b,Tan:2008c} leads to the energy and entropy densities at the production rates of~\cite{Fujii:2018}
\begin{align}\label{eq:response_energy-entropy}
T\dot\S(t) = \dot\E(t) - \frac{\C_\eq[a(t)]}{\Omega_{d-1}ma^{d-1}(t)}\,\dot{a}(t)
= \frac{\C(t)-\C_\eq[a(t)]}{\Omega_{d-1}ma^{d-1}(t)}\,\dot{a}(t).
\end{align}
Therefore, by measuring the contact, energy, or entropy density under the periodic modulation of the scattering length, it is possible in principle to extract the contact--contact response function and thus the bulk viscosity spectral function via Eq.~(\ref{eq:bulk_contact-contact}).

\section{Shear viscosity}\label{sec:shear}
\subsection{Formulation}
The shear viscosity spectral function in Eq.~(\ref{eq:viscosities}) is related to the off-diagonal element of the stress tensor operator via
\begin{align}
\eta(\omega) = \frac{\Im[\chi_{xy,xy}(\omega)]}{\omega},
\end{align}
where the stress--stress response function has the quantum virial expansion of
\begin{align}
\chi_{xy,xy}(\omega)
= \sum_{N=0}^\infty\frac{z^N}{Z}\underbrace{\frac{i}{L^d}\int_0^\infty\!dt\,e^{i(\omega+i0^+)t}
\tr_N[e^{-\beta\hat{H}+i\hat{H}t}\hat\Pi_{xy}e^{-i\hat{H}t}\hat\Pi_{xy}
- e^{-\beta\hat{H}-i\hat{H}t}\hat\Pi_{xy}e^{i\hat{H}t}\hat\Pi_{xy}]}_{\chi_{xy,xy}^{(N)}(\omega)}.
\end{align}
The off-diagonal element of the stress tensor operator (\ref{eq:stress}) then reads
\begin{align}
\hat\Pi_{xy} = -\sum_\sigma\int\!d\x\,
\hat\psi_\sigma^\+(\x)\frac{\d_x\d_y}{m}\hat\psi_\sigma(\x)
= \sum_{N=0}^\infty\underbrace{\openone_N\hat\Pi_{xy}\openone_N}_{\hat\Pi_{xy}^{(N)}}
\end{align}
and its decomposition into one-body and two-body sectors is
\begin{align}
\hat\Pi_{xy}^{(1)} = \sum_\sigma\sum_\p|\sigma\p\>\frac{p_xp_y}{m}\<\sigma\p|
\end{align}
and
\begin{align}
\hat\Pi_{xy}^{(2)} = \sum_{\sigma,\tau}\sum_{\p,\q}
|\sigma\p,\tau\q\>\frac{p_xp_y+q_xq_y}{m}\<\sigma\p,\tau\q|,
\end{align}
respectively.

The stress--stress response function at $O(z)$ is simply provided by
\begin{align}
\chi_{xy,xy}^{(1)}(\omega) = -\frac1{L^d}\sum_{\sigma,\sigma'}\sum_{\p,\p'}
\frac{e^{-\beta\frac{\p^2}{2m}}-e^{-\beta\frac{\p'^2}{2m}}}
{\frac{\p^2}{2m}-\frac{\p'^2}{2m}+\omega+i0^+}
\<\sigma\p|\hat\Pi_{xy}^{(1)}|\sigma'\p'\>\<\sigma'\p'|\hat\Pi_{xy}^{(1)}|\sigma\p\>
\end{align}
and its contribution to the shear viscosity spectral function is thus
\begin{align}\label{eq:shear_one-body}
\eta_1(\omega) = \frac{\Im[\chi_{xy,xy}^{(1)}(\omega)]}{\omega}
= \frac{2\beta}{L^d}\,\pi\delta(\omega)\sum_\p
e^{-\beta\frac{\p^2}{2m}}\left(\frac{p_xp_y}{m}\right)^2.
\end{align}
It is also straightforward to evaluate the stress--stress response function at $O(z^2)$,
\begin{align}
\chi_{xy,xy}^{(2)}(\omega)
= \chi_{xy,xy}^{(2)}(\omega)\big|_\free + \chi_{xy,xy}^{(2)}(\omega)\big|_\inter,
\end{align}
with the use of the time evolution operator such as $e^{-i\hat{H}_2t}$ obtained from Eq.~(\ref{eq:boltzmann}).
The free part with only $e^{-i\hat{H}_2t}|_\free$ involved is simply provided by
\begin{align}
\chi_{xy,xy}^{(2)}(\omega)\big|_\free
= -\frac1{L^d}\sum_{\sigma,\tau}\sum_{\p,\q}\sum_{\sigma',\tau'}\sum_{\p',\q'}
\frac{e^{-\beta\frac{\p^2+\q^2}{2m}}-e^{-\beta\frac{\p'^2+\q'^2}{2m}}}
{\frac{\p^2+\q^2}{2m}-\frac{\p'^2+\q'^2}{2m}+\omega+i0^+}
\<\sigma\p,\tau\q|\hat\Pi_{xy}^{(2)}|\sigma'\p',\tau'\q'\>
\<\sigma'\p',\tau'\q'|\hat\Pi_{xy}^{(2)}|\sigma\p,\tau\q\>.
\end{align}
Its contribution to the shear viscosity spectral function is thus
\begin{align}\label{eq:shear_free}
\eta_2(\omega)\big|_\free = \frac{\Im[\chi_{xy,xy}^{(2)}(\omega)|_\free]}{\omega}
= \underbrace{\frac{4\beta}{L^d}\,\pi\delta(\omega)\sum_{\p,\q}
e^{-\beta\frac{\p^2+\q^2}{2m}}\left(\frac{p_xp_y}{m}\right)^2}_{Z_1\eta_1(\omega)}
- \frac\beta{L^d}\,\pi\delta(\omega)\sum_\p
e^{-\beta\frac{\p^2}{m}}\left(\frac{2p_xp_y}{m}\right)^2.
\end{align}
On the other hand, the interacting part with at least one $e^{-i\hat{H}_2t}|_\inter$ involved is found to be
\begin{align}
\chi_{xy,xy}^{(2)}(\omega)\big|_\inter
&= -\frac1{L^{2d}}\sum_{\p,\q}e^{-\beta\frac{(\p+\q)^2}{4m}}
\left(\frac{p_xp_y+q_xq_y}{m}\right)^2\int_{-\infty}^\infty\!\frac{d\eps}{\pi}
\frac{e^{-\beta\frac{(\p-\q)^2}{4m}}-e^{-\beta\eps}}{\frac{(\p-\q)^2}{4m}-\eps+\omega+i0^+}
\Im\!\left[\frac{\T_2(\eps-i0^+)}{(\eps-\frac{(\p-\q)^2}{4m}-i0^+)^2}\right] \notag\\
&\quad - \frac1{L^{2d}}\sum_{\p,\q}e^{-\beta\frac{(\p+\q)^2}{4m}}
\left(\frac{p_xp_y+q_xq_y}{m}\right)^2\int_{-\infty}^\infty\!\frac{d\eps}{\pi}
\frac{e^{-\beta\eps}-e^{-\beta\frac{(\p-\q)^2}{4m}}}{\eps-\frac{(\p-\q)^2}{4m}+\omega+i0^+}
\Im\!\left[\frac{\T_2(\eps-i0^+)}{(\eps-\frac{(\p-\q)^2}{4m}-i0^+)^2}\right] \notag\\
&\quad - \frac1{L^{3d}}\sum_{\p,\q}\sum_{\p',\q'}e^{-\beta\frac{(\p+\q)^2}{4m}}
\delta_{\p+\q,\p'+\q'}\frac{p_xp_y+q_xq_y}{m}\frac{p'_xp'_y+q'_xq'_y}{m}
\int_{-\infty}^\infty\!\frac{d\eps}{\pi}\int_{-\infty}^\infty\!\frac{d\eps'}{\pi}
\frac{e^{-\beta\eps}-e^{-\beta\eps'}}{\eps-\eps'+\omega+i0^+} \notag\\
&\qquad \times 
\Im\!\left[\frac{\T_2(\eps-i0^+)}{(\eps-\frac{(\p-\q)^2}{4m}-i0^+)(\eps-\frac{(\p'-\q')^2}{4m}-i0^+)}\right]
\Im\!\left[\frac{\T_2(\eps'-i0^+)}{(\eps'-\frac{(\p'-\q')^2}{4m}-i0^+)(\eps'-\frac{(\p-\q)^2}{4m}-i0^+)}\right].
\end{align}
Its contribution to the shear viscosity spectral function is thus
\begin{align}
\eta_2(\omega)\big|_\inter &= \frac{\Im[\chi_{xy,xy}^{(2)}(\omega)|_\inter]}{\omega} \\
&= \frac1{L^{2d}}\sum_{\K,\k}e^{-\beta\frac{\K^2}{4m}-\beta\frac{\k^2}{m}}
\left[\left(\frac{K_xK_y}{2m}\right)^2+\left(\frac{2k_xk_y}{m}\right)^2\right]
\frac{1-e^{-\beta\omega}}{\omega}
\Im\!\left[\frac{\T_2(\frac{\k^2}{m}+\omega-i0^+)}{(\omega-i0^+)^2}\right] \notag\\
&\quad + \frac1{L^{2d}}\sum_{\K,\k}e^{-\beta\frac{\K^2}{4m}-\beta\frac{\k^2}{m}}
\left[\left(\frac{K_xK_y}{2m}\right)^2+\left(\frac{2k_xk_y}{m}\right)^2\right]
\frac{e^{\beta\omega}-1}{\omega}
\Im\!\left[\frac{\T_2(\frac{\k^2}{m}-\omega-i0^+)}{(-\omega-i0^+)^2}\right] \notag\\
&\quad + \frac1{L^{3d}}\sum_\K\sum_{\k,\k'}e^{-\beta\frac{\K^2}{4m}}
\left(\frac{K_xK_y}{2m}\right)^2\frac{1-e^{-\beta\omega}}{\omega}
\int_{-\infty}^\infty\!\frac{d\eps}{\pi}\,e^{-\beta\eps} \notag\\
&\qquad \times
\Im\!\left[\frac{\T_2(\eps-i0^+)}{(\eps-\frac{\k^2}{m}-i0^+)(\eps-\frac{\k'^2}{m}-i0^+)}\right]
\Im\!\left[\frac{\T_2(\eps+\omega-i0^+)}{(\eps+\omega-\frac{\k'^2}{m}-i0^+)(\eps+\omega-\frac{\k^2}{m}-i0^+)}\right],
\end{align}
where $\K=\p+\q$ and $\k^{(\prime)}=(\p^{(\prime)}-\q^{(\prime)})/2$ are the center-of-mass and relative momenta, respectively, and the first two terms correspond to the self-energy and Maki--Thompson diagrams and the last term corresponds to the Aslamazov--Larkin diagram.
The last term can be further simplified by utilizing the identity of
\begin{align}\label{eq:derivation}
& \frac1{L^{2d}}\sum_{\k,\k'}
\Im\!\left[\frac{\T_2(\eps-i0^+)}{(\eps-\frac{\k^2}{m}-i0^+)(\eps-\frac{\k'^2}{m}-i0^+)}\right]
\Im\!\left[\frac{\T_2(\eps+\omega-i0^+)}{(\eps+\omega-\frac{\k'^2}{m}-i0^+)(\eps+\omega-\frac{\k^2}{m}-i0^+)}\right] \notag\\
&= -\frac1{L^d}\sum_\k\pi\delta(\eps-\tfrac{\k^2}{m})
\Im\!\left[\frac{\T_2(\eps+\omega-i0^+)}{(\omega-i0^+)^2}\right]
- \frac1{L^d}\sum_\k\pi\delta(\eps+\omega-\tfrac{\k^2}{m})
\Im\!\left[\frac{\T_2(\eps-i0^+)}{(-\omega-i0^+)^2}\right] \notag\\
&\quad + \frac1{L^d}\sum_\k\pi\delta(\omega)
\Im\!\left[\frac{\T_2(\eps-i0^+)}{(\eps-\frac{\k^2}{m}-i0^+)^2}\right],
\end{align}
so that we obtain
\begin{align}\label{eq:shear_int}
\eta_2(\omega)\big|_\inter
&= \frac1{L^{2d}}\sum_{\K,\k}e^{-\beta\frac{\K^2}{4m}-\beta\frac{\k^2}{m}}
\left(\frac{2k_xk_y}{m}\right)^2\frac{1-e^{-\beta\omega}}{\omega}
\Im\!\left[\frac{\T_2(\frac{\k^2}{m}+\omega-i0^+)}{(\omega-i0^+)^2}\right] \notag\\
&\quad + \frac1{L^{2d}}\sum_{\K,\k}e^{-\beta\frac{\K^2}{4m}-\beta\frac{\k^2}{m}}
\left(\frac{2k_xk_y}{m}\right)^2\frac{e^{\beta\omega}-1}{\omega}
\Im\!\left[\frac{\T_2(\frac{\k^2}{m}-\omega-i0^+)}{(-\omega-i0^+)^2}\right] \notag\\
&\quad + \frac\beta{L^{2d}}\sum_{\K,\k}e^{-\beta\frac{\K^2}{4m}}
\left(\frac{K_xK_y}{2m}\right)^2\pi\delta(\omega)
\int_{-\infty}^\infty\!\frac{d\eps}{\pi}\,e^{-\beta\eps}
\Im\!\left[\frac{\T_2(\eps-i0^+)}{(\eps-\frac{\k^2}{m}-i0^+)^2}\right].
\end{align}
Finally, by combining Eqs.~(\ref{eq:shear_one-body}), (\ref{eq:shear_free}), and (\ref{eq:shear_int}), taking the thermodynamic limit, and performing the resulting momentum integrations, we find that the shear viscosity spectral function is provided by
\begin{align}
\eta(\omega) &= \frac{z}{Z}\,\eta_1(\omega)
+ \frac{z^2}{Z}\left[\eta_2(\omega)\big|_\free
+ \eta_2(\omega)\big|_\inter\right] + O(z^3) \\\label{eq:shear}
&= p\,\pi\delta(\omega)
+ \frac{z^2}{\lambda_T^d}\frac{(m/2\pi)^{d/2}}{\Gamma(2+d/2)}
\frac{1-e^{-\omega/T}}{\omega}\int_0^\infty\!d\eps\,e^{-\eps/T}\eps^{1+d/2}
\Im\!\left[\frac{\T_2(\eps+\omega-i0^+)}{(\omega-i0^+)^2}\right] \notag\\
&\quad + \frac{z^2}{\lambda_T^d}\frac{(m/2\pi)^{d/2}}{\Gamma(2+d/2)}
\frac{e^{\omega/T}-1}{\omega}\int_0^\infty\!d\eps\,e^{-\eps/T}\eps^{1+d/2}
\Im\!\left[\frac{\T_2(\eps-\omega-i0^+)}{(-\omega-i0^+)^2}\right] + O(z^3),
\end{align}
where $p$ in the first term is the pressure from Eq.~(\ref{eq:pressure}) and the integration variable is changed to $\eps=\k^2/m$.

\subsection{Evaluation}
Our concise formula for the shear viscosity spectral function is even under $\omega\to-\omega$ and has the high-frequency tail of
\begin{align}\label{eq:shear_tail}
\lim_{\omega\to\infty}\eta(\omega)
= \frac{\pi\,\C}{(4\pi)^{d/2}\Gamma(2+d/2)|m\omega|^{2-d/2}},
\end{align}
where $\C$ is the contact density from Eq.~(\ref{eq:contact-density}).
This asymptotic form agrees with Refs.~\cite{Enss:2011,Hofmann:2011,Goldberger:2012} for $d=3$ and with Ref.~\cite{Hofmann:2011} for $d=2$.
Also, the integration over $\omega$ leads to
\begin{align}
\int_{-\infty}^\infty\!\frac{d\omega}{\pi}\,\eta(\omega)
= p - \frac{2z^2}{\lambda_T^d}\frac{(m/2\pi)^{d/2}}{\Gamma(2+d/2)}
\int_0^\infty\!d\eps\int_{-\infty}^\infty\!\frac{d\omega}{\pi}\,e^{-\omega/T}\eps^{1+d/2}
\Im\!\left[\frac{\T_2(\omega-i0^+)}{(\omega-\eps-i0^+)^3}\right] + O(z^3).
\end{align}
By performing the integration over $\eps$ with the divergent piece transposed to the left hand side and utilizing the identity of
\begin{align}
\Im[(-m\omega+i0^+)^{d/2-1}\T_2(\omega-i0^+)] = a^{2-d}\Im[\T_2(\omega-i0^+)],
\end{align}
we obtain
\begin{align}\label{eq:shear_sum-rule}
\int_{-\infty}^\infty\!\frac{d\omega}{\pi}
\left[\eta(\omega) - \frac{\pi\,\C}{(4\pi)^{d/2}\Gamma(2+d/2)|m\omega|^{2-d/2}}\right]
= p - \frac\C{(d-2)\Omega_{d-1}ma^{d-2}} \qquad (d>2),
\end{align}
which for $d=3$ is consistent with the sum rule derived in Ref.~\cite{Taylor:2010} up to the factors in front of $\C$ and in Refs.~\cite{Enss:2011,Hofmann:2011}.

After confirming our formula (\ref{eq:shear}) in light of the known constraints, we are ready to evaluate its static limit, $\omega\to0$, where Eq.~(\ref{eq:shear}) is formally ill-defined and thus requires the resummation.
To this end, we write $\pi\delta(\omega)=\Im(\omega-i0^+)^{-1}$ in the first term and regard the remainder as corrections in a geometric series~\cite{Gotze:1972,Enss:2011}.
Consequently, the shear viscosity spectral function is brought into the Drude form of
\begin{align}
\eta(\omega) = \Im\!\left[\frac{p}{\omega-i0^+-\M(\omega)}\right],
\end{align}
where the memory function is found to be
\begin{align}
\M(\omega) &= \frac{z^2}{p\lambda_T^d}\frac{(m/2\pi)^{d/2}}{\Gamma(2+d/2)}
\frac{1-e^{-\omega/T}}{\omega}\int_0^\infty\!d\eps\,e^{-\eps/T}\eps^{1+d/2}\T_2(\eps+\omega-i0^+) \notag\\
&\quad - \frac{z^2}{p\lambda_T^d}\frac{(m/2\pi)^{d/2}}{\Gamma(2+d/2)}
\frac{e^{\omega/T}-1}{\omega}\int_0^\infty\!d\eps\,e^{-\eps/T}\eps^{1+d/2}\T_2(\eps-\omega+i0^+)
+ O(z^2).
\end{align}
Because of $\Re[\M(0)]=0$, the static limit is now provided by $\eta\equiv\lim_{\omega\to0}\eta(\omega)=p/\Im[\M(0)]$ and the substitution of Eqs.~(\ref{eq:pressure}) and (\ref{eq:T-matrix}) therein leads to
\begin{align}\label{eq:shear_static}
\frac1{\lambda_T^d\eta} =
\begin{cases}\displaystyle\,
\frac\pi2\int_0^\infty\!d\tilde\eps\,e^{-\tilde\eps}
\frac{\tilde\eps^2}{\ln^2(\tilde{a}^2\tilde\eps)+\pi^2} + O(z) &\quad (d=2),
\medskip\\\displaystyle\,
\frac{4\sqrt2}{15\pi}\int_0^\infty\!d\tilde\eps\,e^{-\tilde\eps}
\frac{\tilde\eps^3}{\tilde\eps+\frac1{\tilde{a}^2}} + O(z) &\quad (d=3).
\end{cases}
\end{align}
Here the integration variable is changed to $\tilde\eps=\eps/T$ and the dimensionless scattering length is introduced via $\tilde{a}\equiv\sqrt{mT}a=\sqrt{2\pi}a/\lambda_T$.
Our results agree with the static shear viscosity derived from the kinetic theory both for $d=3$~\cite{Massignan:2005,Bruun:2005} and for $d=2$~\cite{Bruun:2012,Schafer:2012}.
Although it is a smooth function of $\lambda_T/a>0$ for $d=2$, we find that the static shear viscosity for $d=3$ exhibits the weak non-analyticity starting with $\lambda_T^3\eta\sim(\ln\tilde{a}^2)/\tilde{a}^6$ in the unitarity limit.

\acknowledgments
The author thanks Thomas Sch\"afer for the valuable discussions.
He also thanks Tilman Enss and Johannes Hofmann for the correspondences regarding their recent preprints~\cite{Enss:2019,Hofmann:2019}, which overlap with this work.
This work was supported by JSPS KAKENHI Grant Nos.~JP15K17727 and JP15H05855.

\appendix\section{Bulk viscosity formula at $O(z^3)$}\label{appendix}
Let us derive the formula for the contact--contact response function at $O(z^3)$, which is to evaluate
\begin{align}
\chi_{CC}^{(3)}(\omega)
= \frac{i}{L^d}\int_0^\infty\!dt\,e^{i(\omega+i0^+)t}
\tr_3[e^{-\beta\hat{H}+i\hat{H}t}\hat{C}e^{-i\hat{H}t}\hat{C}
- e^{-\beta\hat{H}-i\hat{H}t}\hat{C}e^{i\hat{H}t}\hat{C}].
\end{align}
The three-body sector is spanned by $|\sigma_1\p_1,\sigma_2\p_2,\sigma_3\p_3\>$, which is further decomposed into four subsectors where $\{\sigma_1,\sigma_2,\sigma_3\}=\{\up,\up,\up\},\,\{\up,\up,\down\},\,\{\up,\down,\down\},\,\{\down,\down,\down\}$.
The first and last subsectors do not involve the interaction and thus do not contribute to the contact--contact response function.
Because the second and third subsectors contribute identically, it is sufficient to consider only the second subsector.
By introducing the abbreviated notation,
\begin{align}
|\p_1,\p_2;\q\> \equiv \sqrt3\,|\up\p_1,\up\p_2,\down\q\>
= \frac1{\sqrt2}\,\hat\psi_{\up\p_1}^\+\hat\psi_{\up\p_2}^\+\hat\psi_{\down\q}^\+|0\>,
\end{align}
and utilizing the identity operator,
\begin{align}
\openone_{\up\up\down} = \sum_{\p_1,\p_2,\q}|\p_1,\p_2;\q\>\<\p_1,\p_2;\q|,
\end{align}
the projection of the Hamiltonian~(\ref{eq:hamiltonian}) and the contact operator~(\ref{eq:contact}) onto this subsector leads to
\begin{align}
\hat{H}_{\up\up\down} &\equiv \openone_{\up\up\down}\hat{H}\openone_{\up\up\down} \\
&= \underbrace{\sum_{\p_1,\p_2,\q}
|\p_1,\p_2;\q\>\frac{\p_1^2+\p_2^2+\q^2}{2m}\<\p_1,\p_2;\q|}_{\hat{T}_{\up\up\down}}
+ \underbrace{\sum_{\p_1,\p_2,\q}\sum_{\p'_1,\p_2';\q'}|\p_1,\p_2;\q\>
2\delta_{\p_1,\p_1'}\delta_{\p_2+\q,\p_2'+\q'}\frac{g}{L^d}\<\p'_1,\p_2';\q'|}_{\hat{V}_{\up\up\down}}
\end{align}
and
\begin{align}
\hat{C}_{\up\up\down} \equiv \openone_{\up\up\down}\hat{C}\openone_{\up\up\down}
= \sum_{\p_1,\p_2,\q}\sum_{\p'_1,\p_2';\q'}|\p_1,\p_2;\q\>
2\delta_{\p_1,\p_1'}\delta_{\p_2+\q,\p_2'+\q'}\frac{(mg)^2}{L^d}\<\p'_1,\p_2';\q'|.
\end{align}

In order to evaluate the corresponding Boltzmann operator, we write the Hamiltonian as a sum of kinetic and potential energy operators and utilize the identity of
\begin{align}
e^{-\beta\hat{H}_{\up\up\down}} &= \int_{-\infty}^\infty\!\frac{dE}{\pi}\,e^{-\beta E}
\Im\!\left[\frac1{E-\hat{H}_{\up\up\down}-i0^+}\right] \\
&= \int_{-\infty}^\infty\!\frac{dE}{\pi}\,e^{-\beta E}
\Im\!\left[\frac1{E-\hat{T}_{\up\up\down}-i0^+}
+ \sum_{n=0}^\infty\frac1{E-\hat{T}_{\up\up\down}-i0^+}\hat{V}_{\up\up\down}
\left(\frac1{E-\hat{T}_{\up\up\down}-i0^+}\hat{V}_{\up\up\down}\right)^n
\frac1{E-\hat{T}_{\up\up\down}-i0^+}\right].
\end{align}
Because the summation over $n$ can be performed according to
\begin{align}
& \sum_{n=0}^\infty\hat{V}_{\up\up\down}
\left(\frac1{E-\hat{T}_{\up\up\down}-i0^+}\hat{V}_{\up\up\down}\right)^n \notag\\
&= \frac1{L^d}\sum_{\p_1,\p_2,\q}\sum_{\p'_1,\p_2',\q'}
|\p_1,\p_2;\q\>2\delta_{\p_1\p'_1}\delta_{\p_2+\q,\p_2'+\q'}
\T_2(E-\tfrac{\p_1^2}{2m}-\tfrac{(\p_2+\q)^2}{4m}-i0^+)
\<\p'_1,\p_2';\q'| \notag\\
&\quad + \frac1{L^{2d}}\sum_{\p_1,\p_2,\q}\sum_{\p'_1,\p_2',\q'}
|\p_1,\p_2;\q\>2\delta_{\p_1+\p_2+\q,\p'_1+\p_2'+\q'}
\T_2(E-\tfrac{\p_1^2}{2m}-\tfrac{(\p_2+\q)^2}{4m}-i0^+) \notag\\
&\qquad \times \tilde\T_3(E-i0^+,\p_1+\p_2+\q;\p_1,\p_1')
\T_2(E-\tfrac{\p_1'^2}{2m}-\tfrac{(\p_2'+\q')^2}{4m}-i0^+)
\<\p'_1,\p_2';\q'|,
\end{align}
the Boltzmann operator is found to be
\begin{align}
& e^{-\beta\hat{H}_{\up\up\down}}
= \sum_{\p_1,\p_2,\q}|\p_1,\p_2;\q\>
e^{-\beta\frac{\p_1^2+\p_2^2+\q^2}{2m}}\<\p_1,\p_2;\q| \notag\\
&+ \frac1{L^d}\sum_{\p_1,\p_2,\q}\sum_{\p'_1,\p_2',\q'}
|\p_1,\p_2;\q\>2\delta_{\p_1\p'_1}\delta_{\p_2+\q,\p_2'+\q'}\<\p'_1,\p_2';\q'| \notag\\
&\quad \times \int_{-\infty}^\infty\!\frac{dE}{\pi}\,e^{-\beta E}
\Im\!\left[\frac{\T_2(E-\frac{\p_1^2}{2m}-\frac{(\p_2+\q)^2}{4m}-i0^+)}
{(E-\frac{\p_1^2+\p_2^2+\q^2}{2m}-i0^+)(E-\frac{\p_1'^2+\p_2'^2+\q'^2}{2m}-i0^+)}\right] \notag\\
&+ \frac1{L^{2d}}\sum_{\p_1,\p_2,\q}\sum_{\p'_1,\p_2',\q'}
|\p_1,\p_2;\q\>2\delta_{\p_1+\p_2+\q,\p'_1+\p_2'+\q'}\<\p'_1,\p_2';\q'| \notag\\
&\quad \times \int_{-\infty}^\infty\!\frac{dE}{\pi}\,e^{-\beta E}
\Im\!\left[\frac{\T_2(E-\frac{\p_1^2}{2m}-\frac{(\p_2+\q)^2}{4m}-i0^+)
\tilde\T_3(E-i0^+,\p_1+\p_2+\q;\p_1,\p_1')
\T_2(E-\frac{\p_1'^2}{2m}-\frac{(\p_2'+\q')^2}{4m}-i0^+)}
{(E-\frac{\p_1^2+\p_2^2+\q^2}{2m}-i0^+)(E-\frac{\p_1'^2+\p_2'^2+\q'^2}{2m}-i0^+)}\right].
\end{align}
Here an infinite series,
\begin{align}
& \tilde\T_3(E-i0^+,\K;\p_1,\p_1')
\equiv \frac{-1}{E-\frac{\p_1^2+\p_1'^2+(\K-\p_1-\p_1')^2}{2m}-i0^+} \notag\\
&+ \frac1{L^d}\sum_{\p_1''}\frac{-1}{E-\frac{\p_1^2+\p_1''^2+(\K-\p_1-\p_1'')^2}{2m}-i0^+}
\T_2(E-\tfrac{\p_1''^2}{2m}-\tfrac{(\K-\p_1'')^2}{4m}-i0^+)
\frac{-1}{E-\frac{\p_1''^2+\p_1'^2+(\K-\p_1''-\p_1')^2}{2m}-i0^+}
+ \cdots,
\end{align}
is the three-body scattering $T$-matrix between atom and dimer from incoming $(\p_1,\K-\p_1)$ to outgoing $(\p_1',\K-\p_1')$ momenta and solves the integral equation of~\cite{Braaten:2006,Barth:2015}
\begin{align}
& \tilde\T_3(E-i0^+,\K;\p_1,\p_1') \notag\\
&= -\frac1{E-\frac{\p_1^2+\p_1'^2+(\K-\p_1-\p_1')^2}{2m}-i0^+}
- \frac1{L^d}\sum_{\p_1''}
\frac{\T_2(E-\frac{\p_1''^2}{2m}-\frac{(\K-\p_1'')^2}{4m}-i0^+)}
{E-\frac{\p_1^2+\p_1''^2+(\K-\p_1-\p_1'')^2}{2m}-i0^+}
\tilde\T_3(E-i0^+,\K;\p_1'',\p_1').
\end{align}

With the use of the above Boltzmann operator valid for any complex $\beta$ as well as the identity of Eq.~(\ref{eq:identity}), we obtain
\begin{align}
& \tr_{\up\up\down}[e^{-\beta\hat{H}_{\up\up\down}+i\hat{H}_{\up\up\down}t}
\hat{C}_{\up\up\down}e^{-i\hat{H}_{\up\up\down}t}\hat{C}_{\up\up\down}] \notag\\
&= \sum_{\K,\k}e^{-\beta\frac{\K^2}{6m}}\int_{-\infty}^\infty\!\frac{d\eps}{\pi}
\int_{-\infty}^\infty\!\frac{d\eps'}{\pi}\,e^{-\beta\eps+i\eps t-i\eps't}
\Im\!\left[m^2\T_2(\eps-\tfrac{3\k^2}{4m}-i0^+)\right]
\Im\!\left[m^2\T_2(\eps'-\tfrac{3\k^2}{4m}-i0^+)\right] \notag\\
&\quad + \frac1{L^d}\sum_{\K,\k}
e^{-\beta\frac{\K^2}{6m}}\int_{-\infty}^\infty\!\frac{d\eps}{\pi}
\int_{-\infty}^\infty\!\frac{d\eps'}{\pi}\,e^{-\beta\eps+i\eps t-i\eps't}
\Im\!\left[m^2\T_2(\eps-\tfrac{3\k^2}{4m}-i0^+)\right] \notag\\
&\qquad \times \Im\!\left[m^2\T_2(\eps'-\tfrac{3\k^2}{4m}-i0^+)
\tilde\T_3(\eps'+\tfrac{\K^2}{6m}-i0^+,\K;\tfrac\K3+\k,\tfrac\K3+\k)
\T_2(\eps'-\tfrac{3\k^2}{4m}-i0^+)\right] \notag\\
&\quad + \frac1{L^d}\sum_{\K,\k}
e^{-\beta\frac{\K^2}{6m}}\int_{-\infty}^\infty\!\frac{d\eps}{\pi}
\int_{-\infty}^\infty\!\frac{d\eps'}{\pi}\,e^{-\beta\eps+i\eps t-i\eps't}
\Im\!\left[m^2\T_2(\eps'-\tfrac{3\k^2}{4m}-i0^+)\right] \notag\\
&\qquad \times \Im\!\left[m^2\T_2(\eps-\tfrac{3\k^2}{4m}-i0^+)
\tilde\T_3(\eps+\tfrac{\K^2}{6m}-i0^+,\K;\tfrac\K3+\k,\tfrac\K3+\k)
\T_2(\eps-\tfrac{3\k^2}{4m}-i0^+)\right] \notag\\
&\quad + \frac1{L^{2d}}\sum_{\K,\k,\k'}
e^{-\beta\frac{\K^2}{6m}}\int_{-\infty}^\infty\!\frac{d\eps}{\pi}
\int_{-\infty}^\infty\!\frac{d\eps'}{\pi}\,e^{-\beta\eps+i\eps t-i\eps't} \notag\\
&\qquad \times \Im\!\left[m^2\T_2(\eps-\tfrac{3\k^2}{4m}-i0^+)
\tilde\T_3(\eps+\tfrac{\K^2}{6m}-i0^+,\K;\tfrac\K3+\k,\tfrac\K3+\k')
\T_2(\eps-\tfrac{3\k'^2}{4m}-i0^+)\right] \notag\\
&\qquad \times \Im\!\left[m^2\T_2(\eps'-\tfrac{3\k'^2}{4m}-i0^+)
\tilde\T_3(\eps'+\tfrac{\K^2}{6m}-i0^+,\K;\tfrac\K3+\k',\tfrac\K3+\k)
\T_2(\eps'-\tfrac{3\k^2}{4m}-i0^+)\right],
\end{align}
where $\K=\p_1+\p_2+\q$ and $\k^{(\prime)}=\p_1^{(\prime)}-\K/3$ are the center-of-mass and relative momenta, respectively, and the integration variables are changed to $\eps^{(\prime)}=E^{(\prime)}-\K^2/6m$.
We also note that $\T_3(\eps-i0^+;\k,\k')\equiv\tilde\T_3(\eps+\frac{\K^2}{6m}-i0^+,\K;\frac\K3+\k,\frac\K3+\k')$ is the three-body scattering $T$-matrix in the center-of-mass frame and is actually independent of $\K$.
Consequently, the contact--contact response function at $O(z^3)$ is found to be
\begin{align}
\chi_{CC}^{(3)}(\omega) &= -\frac2{L^d}\sum_{\K,\k}
e^{-\beta\frac{\K^2}{6m}}\int_{-\infty}^\infty\!\frac{d\eps}{\pi}
\int_{-\infty}^\infty\!\frac{d\eps'}{\pi}
\frac{e^{-\beta\eps}-e^{-\beta\eps'}}{\omega+\eps-\eps'+i0^+}
\Im\!\left[m^2\T_2(\eps-\tfrac{3\k^2}{4m}-i0^+)\right]
\Im\!\left[m^2\T_2(\eps'-\tfrac{3\k^2}{4m}-i0^+)\right] \notag\\
&\quad - \frac2{L^{2d}}\sum_{\K,\k}
e^{-\beta\frac{\K^2}{6m}}\int_{-\infty}^\infty\!\frac{d\eps}{\pi}
\int_{-\infty}^\infty\!\frac{d\eps'}{\pi}
\frac{e^{-\beta\eps}-e^{-\beta\eps'}}{\omega+\eps-\eps'+i0^+}
\Im\!\left[m^2\T_2(\eps-\tfrac{3\k^2}{4m}-i0^+)\right] \notag\\
&\qquad \times \Im\!\left[m^2\T_2(\eps'-\tfrac{3\k^2}{4m}-i0^+)
\T_3(\eps'-i0^+;\k,\k)\T_2(\eps'-\tfrac{3\k^2}{4m}-i0^+)\right] \notag\\
&\quad - \frac2{L^{2d}}\sum_{\K,\k}
e^{-\beta\frac{\K^2}{6m}}\int_{-\infty}^\infty\!\frac{d\eps}{\pi}
\int_{-\infty}^\infty\!\frac{d\eps'}{\pi}
\frac{e^{-\beta\eps}-e^{-\beta\eps'}}{\omega+\eps-\eps'+i0^+}
\Im\!\left[m^2\T_2(\eps'-\tfrac{3\k^2}{4m}-i0^+)\right] \notag\\
&\qquad \times \Im\!\left[m^2\T_2(\eps-\tfrac{3\k^2}{4m}-i0^+)
\T_3(\eps-i0^+;\k,\k)\T_2(\eps-\tfrac{3\k^2}{4m}-i0^+)\right] \notag\\
&\quad - \frac2{L^{3d}}\sum_{\K,\k,\k'}
e^{-\beta\frac{\K^2}{6m}}\int_{-\infty}^\infty\!\frac{d\eps}{\pi}
\int_{-\infty}^\infty\!\frac{d\eps'}{\pi}
\frac{e^{-\beta\eps}-e^{-\beta\eps'}}{\omega+\eps-\eps'+i0^+} \notag\\
&\qquad \times \Im\!\left[m^2\T_2(\eps-\tfrac{3\k^2}{4m}-i0^+)
\T_3(\eps-i0^+;\k,\k')\T_2(\eps-\tfrac{3\k'^2}{4m}-i0^+)\right] \notag\\
&\qquad \times \Im\!\left[m^2\T_2(\eps'-\tfrac{3\k'^2}{4m}-i0^+)
\T_3(\eps'-i0^+;\k',\k)\T_2(\eps'-\tfrac{3\k^2}{4m}-i0^+)\right],
\end{align}
where the overall factor of 2 accounts for the existence of two interacting subsectors and the first term corresponds to the disconnected diagram, the second and third terms correspond to the self-energy diagrams, and the last term corresponds to the Aslamazov--Larkin diagram.
Finally, by combining Eq.~(\ref{eq:bulk_two-body}), taking the thermodynamic limit, and performing the resulting momentum integrations, we find that the bulk viscosity spectral function is provided by
\begin{align}
\zeta(\omega) &= \frac1{(d\,\Omega_{d-1}ma^{d-2})^2\,\omega}
\Im\!\left[\frac{z^2}{Z}\,\chi_{CC}^{(2)}(\omega)
+ \frac{z^3}{Z}\,\chi_{CC}^{(3)}(\omega) + O(z^4)\right] \\\label{eq:bulk_three-body}
&= \frac{2^{d/2}z^2}{(d\,\Omega_{d-1}a^{d-2})^2\lambda_T^d}
\frac{1-e^{-\omega/T}}{\omega}\int_{-\infty}^\infty\!\frac{d\eps}{\pi}\,
e^{-\eps/T}\Im[m\T_2(\eps-i0^+)]\Im[m\T_2(\eps+\omega-i0^+)] \notag\\
&\quad + \frac{2\cdot3^{d/2}z^3}{(d\,\Omega_{d-1}a^{d-2})^2\lambda_T^d}
\frac{1-e^{-\omega/T}}{\omega}\int\!\frac{d\k}{(2\pi)^d}
\int_{-\infty}^\infty\!\frac{d\eps}{\pi}\,e^{-\eps/T}
\Im\!\left[m\T_2(\eps-\tfrac{3\k^2}{4m}-i0^+)\right] \notag\\
&\qquad \times \Im\!\left[m\T_2(\eps+\omega-\tfrac{3\k^2}{4m}-i0^+)
\T_3(\eps+\omega-i0^+;\k,\k)\T_2(\eps+\omega-\tfrac{3\k^2}{4m}-i0^+)\right] \notag\\
&\quad + \frac{2\cdot3^{d/2}z^3}{(d\,\Omega_{d-1}a^{d-2})^2\lambda_T^d}
\frac{1-e^{-\omega/T}}{\omega}\int\!\frac{d\k}{(2\pi)^d}
\int_{-\infty}^\infty\!\frac{d\eps}{\pi}\,e^{-\eps/T}
\Im\!\left[m\T_2(\eps+\omega-\tfrac{3\k^2}{4m}-i0^+)\right] \notag\\
&\qquad \times \Im\!\left[m\T_2(\eps-\tfrac{3\k^2}{4m}-i0^+)
\T_3(\eps-i0^+;\k,\k)\T_2(\eps-\tfrac{3\k^2}{4m}-i0^+)\right] \notag\\
&\quad + \frac{2\cdot3^{d/2}z^3}{(d\,\Omega_{d-1}a^{d-2})^2\lambda_T^d}
\frac{1-e^{-\omega/T}}{\omega}\int\!\frac{d\k}{(2\pi)^d}\int\!\frac{d\k'}{(2\pi)^d}
\int_{-\infty}^\infty\!\frac{d\eps}{\pi}\,e^{-\eps/T} \notag\\
&\qquad \times \Im\!\left[m\T_2(\eps-\tfrac{3\k^2}{4m}-i0^+)
\T_3(\eps-i0^+;\k,\k')\T_2(\eps-\tfrac{3\k'^2}{4m}-i0^+)\right] \notag\\
&\qquad \times \Im\!\left[m\T_2(\eps+\omega-\tfrac{3\k'^2}{4m}-i0^+)
\T_3(\eps+\omega-i0^+;\k',\k)\T_2(\eps+\omega-\tfrac{3\k^2}{4m}-i0^+)\right] + O(z^4),
\end{align}
where the three-body scattering $T$-matrix in the center-of-mass frame solves the integral equation of
\begin{align}
\T_3(\eps-i0^+;\k,\k') = -\frac1{\eps-\frac{\k^2+\k'^2+\k\cdot\k'}{m}-i0^+}
- \int\!\frac{d\k''}{(2\pi)^d}\frac{\T_2(\eps-\frac{3\k''^2}{4m}-i0^+)}
{\eps-\frac{\k^2+\k''^2+\k\cdot\k''}{m}-i0^+}\T_3(\eps-i0^+;\k'',\k').
\end{align}
The higher order correction at $O(z^3)$ may be useful to better understand the static bulk viscosity in the Kubo formalism, which is to be investigated in a future study.

\section{Derivation of Eq.~(\ref{eq:derivation})}
We first decompose the left-hand side of Eq.~(\ref{eq:derivation}) as
\begin{align}
& \mathrm{LHS} \equiv \frac1{L^{2d}}\sum_{\k,\k'}
\Im\!\left[\frac{\T_2(\eps-i0^+)}{(\eps-\frac{\k^2}{m}-i0^+)(\eps-\frac{\k'^2}{m}-i0^+)}\right]
\Im\!\left[\frac{\T_2(\eps+\omega-i0^+)}{(\eps+\omega-\frac{\k'^2}{m}-i0^+)(\eps+\omega-\frac{\k^2}{m}-i0^+)}\right] \\
&= \frac1{L^{2d}}\sum_{\k,\k'}
\Im\!\left[\frac{\T_2(\eps-i0^+)}{2i(\eps-\frac{\k^2}{m}-i0^+)(\eps-\frac{\k'^2}{m}-i0^+)}
\frac{\T_2(\eps+\omega-i0^+)}{(\eps+\omega-\frac{\k'^2}{m}-i0^+)(\eps+\omega-\frac{\k^2}{m}-i0^+)}\right] \notag\\
&\quad - \frac1{L^{2d}}\sum_{\k,\k'}
\Im\!\left[\frac{\T_2(\eps+i0^+)}{2i(\eps-\frac{\k^2}{m}+i0^+)(\eps-\frac{\k'^2}{m}+i0^+)}
\frac{\T_2(\eps+\omega-i0^+)}{(\eps+\omega-\frac{\k'^2}{m}-i0^+)(\eps+\omega-\frac{\k^2}{m}-i0^+)}\right] \\
&= \frac1{L^{2d}}\sum_{\k,\k'}
\Im\!\left[\frac{\T_2(\eps-i0^+)\T_2(\eps+\omega-i0^+)}{2i(\omega-i0^+)^2}
\left[\frac1{\eps-\frac{\k^2}{m}-i0^+} - \frac1{\eps+\omega-\frac{\k^2}{m}-i0^+}\right]
\left[\frac1{\eps-\frac{\k'^2}{m}-i0^+} - \frac1{\eps+\omega-\frac{\k'^2}{m}-i0^+}\right]\right] \notag\\
&\quad - \frac1{L^{2d}}\sum_{\k,\k'}
\Im\!\left[\frac{\T_2(\eps+i0^+)\T_2(\eps+\omega-i0^+)}{2i(\omega-i0^+)^2}
\left[\frac1{\eps-\frac{\k^2}{m}+i0^+} - \frac1{\eps+\omega-\frac{\k^2}{m}-i0^+}\right]
\left[\frac1{\eps-\frac{\k'^2}{m}+i0^+} - \frac1{\eps+\omega-\frac{\k'^2}{m}-i0^+}\right]\right].
\end{align}
Then, the definition of the two-body scattering $T$-matrix,
\begin{align}
\frac1{\T_2(\eps-i0^+)} = \frac1g - \frac1{L^d}\sum_\k\frac1{\eps-\frac{\k^2}{m}-i0^+},
\end{align}
leads to
\begin{align}
\mathrm{LHS}
&= \Im\!\left[\frac{\T_2(\eps+\omega-i0^+) - \T_2(\eps-i0^+)}{2i(\omega-i0^+)^2}
\left[\frac1{\T_2(\eps-i0^+)} - \frac1{\T_2(\eps+\omega-i0^+)}\right]\right] \notag\\
&\quad - \Im\!\left[\frac{\T_2(\eps+\omega-i0^+) - \T_2(\eps+i0^+)}{2i(\omega-i0^+)^2}
\left[\frac1{\T_2(\eps+i0^+)} - \frac1{\T_2(\eps+\omega-i0^+)}\right]\right] \\
&= \Im\!\left[\frac{\T_2(\eps+\omega-i0^+)}{2i(\omega-i0^+)^2}
\left[\frac1{\T_2(\eps-i0^+)} - \frac1{\T_2(\eps+i0^+)}\right]\right] \notag\\
&\quad + \Im\!\left[\frac{\T_2(\eps-i0^+)}{2i(\omega+i0^+)^2}
\left[\frac1{\T_2(\eps+\omega-i0^+)} - \frac1{\T_2(\eps+\omega+i0^+)}\right]\right] \notag\\
&\quad - \Im\!\left[\left[\frac{\T_2(\eps-i0^+)}{2i(\omega-i0^+)^2} - \frac{\T_2(\eps-i0^+)}{2i(\omega+i0^+)^2}\right]
\left[\frac1{\T_2(\eps-i0^+)} - \frac1{\T_2(\eps+\omega-i0^+)}\right]\right].
\end{align}
Finally, with the use of
\begin{align}
\frac1{\T_2(\eps-i0^+)} - \frac1{\T_2(\eps+i0^+)}
= -\frac1{L^d}\sum_\k2\pi i\,\delta(\eps-\tfrac{\k^2}{m})
\end{align}
and
\begin{align}
\frac1{(\omega-i0^+)^2} - \frac1{(\omega+i0^+)^2}
= -\frac{d}{d\omega}\left[\frac1{\omega-i0^+} - \frac1{\omega+i0^+}\right]
= -\frac{d}{d\omega}[2\pi i\,\delta(\omega)],
\end{align}
we obtain
\begin{align}
\mathrm{LHS}
&= -\frac1{L^d}\sum_\k\pi\delta(\eps-\tfrac{\k^2}{m})
\Im\!\left[\frac{\T_2(\eps+\omega-i0^+)}{(\omega-i0^+)^2}\right]
- \frac1{L^d}\sum_\k\pi\delta(\eps+\omega-\tfrac{\k^2}{m})
\Im\!\left[\frac{\T_2(\eps-i0^+)}{(\omega+i0^+)^2}\right] \notag\\
&\quad - \frac1{L^d}\sum_\k\pi\delta'(\omega)\Im\!\left[\T_2(\eps-i0^+)
\left[\frac1{\eps-\frac{\k^2}{m}-i0^+} - \frac1{\eps+\omega-\frac{\k^2}{m}-i0^+}\right]\right] \\
&= -\frac1{L^d}\sum_\k\pi\delta(\eps-\tfrac{\k^2}{m})
\Im\!\left[\frac{\T_2(\eps+\omega-i0^+)}{(\omega-i0^+)^2}\right]
- \frac1{L^d}\sum_\k\pi\delta(\eps+\omega-\tfrac{\k^2}{m})
\Im\!\left[\frac{\T_2(\eps-i0^+)}{(\omega+i0^+)^2}\right] \notag\\
&\quad - \frac1{L^d}\sum_\k\pi\,\omega\delta'(\omega)
\Im\!\left[\frac{\T_2(\eps-i0^+)}{(\eps-\frac{\k^2}{m}-i0^+)^2}\right],
\end{align}
which is the right-hand side of Eq.~(\ref{eq:derivation}) under $\omega\delta'(\omega)=-\delta(\omega)$.


\begin{thebibliography}{99}

\bibitem{Schafer:2009}
T.~Sch\"afer and D.~Teaney,
``Nearly perfect fluidity: from cold atomic gases to hot quark gluon plasmas,''
\href{https://doi.org/10.1088/0034-4885/72/12/126001}
{Rep.\ Prog.\ Phys.\ \textbf{72}, 126001 (2009)}.

\bibitem{Adams:2012}
A.~Adams, L.~D.~Carr, T.~Sch\"afer, P.~Steinberg, and J.~E.~Thomas,
``Strongly correlated quantum fluids: ultracold quantum gases, quantum chromodynamic plasmas and holographic duality,''
\href{https://doi.org/10.1088/1367-2630/14/11/115009}
{New J.\ Phys.\ \textbf{14}, 115009 (2012)}.

\bibitem{Kovtun:2005}
P.~K.~Kovtun, D.~T.~Son, and A.~O.~Starinets,
``Viscosity in strongly interacting quantum field theories from black hole physics,''
\href{https://doi.org/10.1103/PhysRevLett.94.111601}
{Phys.\ Rev.\ Lett.\ \textbf{94}, 111601 (2005)}.

\bibitem{Brigante:2008a}
M.~Brigante, H.~Liu, R.~C.~Myers, S.~Shenker, and S.~Yaida,
``Viscosity bound violation in higher derivative gravity,''
\href{https://doi.org/10.1103/PhysRevD.77.126006}
{Phys.\ Rev.\ D \textbf{77}, 126006 (2008)}.

\bibitem{Brigante:2008b}
M.~Brigante, H.~Liu, R.~C.~Myers, S.~Shenker, and S.~Yaida,
``Viscosity bound and causality violation,''
\href{https://doi.org/10.1103/PhysRevLett.100.191601}
{Phys.\ Rev.\ Lett.\ \textbf{100}, 191601 (2008)}.

\bibitem{Kats:2009}
Y.~Kats and P.~Petrov,
``Effect of curvature squared corrections in AdS on the viscosity of the dual gauge theory,''
\href{https://doi.org/10.1088/1126-6708/2009/01/044}
{J.\ High Energy Phys.\ \textbf{01} (2009) 044}.

\bibitem{Buchel:2009}
A.~Buchel, R.~C.~Myers, and A.~Sinha,
``Beyond $\eta/s=1/4\pi$,''
\href{https://doi.org/10.1088/1126-6708/2009/03/084}
{J.\ High Eenergy Phys.\ \textbf{03} (2009) 084}.

\bibitem{Chin:2010}
C.~Chin, R.~Grimm, P.~Julienne, and E.~Tiesinga,
``Feshbach resonances in ultracold gases,''
\href{https://doi.org/10.1103/RevModPhys.82.1225}
{Rev.\ Mod.\ Phys.\ \textbf{82}, 1225-1286 (2010)}.

\bibitem{Cao:2011a}
C.~Cao, E.~Elliott, J.~Joseph, H.~Wu, J.~Petricka, T.~Sch\"afer, and J.~E.~Thomas,
``Universal quantum viscosity in a unitary Fermi gas,''
\href{https://doi.org/10.1126/science.1195219}
{Science \textbf{331},58-61 (2011)}.

\bibitem{Cao:2011b}
C.~Cao, E.~Elliott, H.~Wu, and J.~E.~Thomas,
``Searching for perfect fluids: quantum viscosity in a universal Fermi gas,''
\href{https://doi.org/10.1088/1367-2630/13/7/075007}
{New J.\ Phys.\ \textbf{13}, 075007 (2011)}.

\bibitem{Elliott:2014b}
E.~Elliott, J.~A.~Joseph, and J.~E.~Thomas,
``Anomalous minimum in the shear viscosity of a Fermi gas,''
\href{https://doi.org/10.1103/PhysRevLett.113.020406}
{Phys.\ Rev.\ Lett.\ \textbf{113}, 020406 (2014)}.

\bibitem{Son:2007}
D.~T.~Son,
``Vanishing bulk viscosities and conformal invariance of the unitary Fermi gas,''
\href{https://doi.org/10.1103/PhysRevLett.98.020604}
{Phys.\ Rev.\ Lett.\ \textbf{98}, 020604 (2007)}.

\bibitem{Elliott:2014a}
E.~Elliott, J.~A.~Joseph, and J.~E.~Thomas,
``Observation of conformal symmetry breaking and scale invariance in expanding Fermi gases,''
\href{https://doi.org/10.1103/PhysRevLett.112.040405}
{Phys.\ Rev.\ Lett.\ \textbf{112}, 040405 (2014)}.

\bibitem{Vogt:2012}
E.~Vogt, M.~Feld, B.~Fr\"ohlich, D.~Pertot, M.~Koschorreck, and M.~K\"ohl,
``Scale invariance and viscosity of a two-dimensional Fermi gas,''
\href{https://doi.org/10.1103/PhysRevLett.108.070404}
{Phys.\ Rev.\ Lett.\ \textbf{108}, 070404 (2012)}.

\bibitem{Taylor:2010}
E.~Taylor and M.~Randeria,
``Viscosity of strongly interacting quantum fluids: Spectral functions and sum rules,''
\href{https://doi.org/10.1103/PhysRevA.81.053610}
{Phys.\ Rev.\ A \textbf{81}, 053610 (2010)}.

\bibitem{Enss:2011}
T.~Enss, R.~Haussmann, and W.~Zwerger,
``Viscosity and scale invariance in the unitary Fermi gas,''
\href{https://doi.org/10.1016/j.aop.2010.10.002}
{Ann.\ Phys.\ \textbf{326}, 770-796 (2011)}.

\bibitem{Hofmann:2011}
J.~Hofmann,
``Current response, structure factor and hydrodynamic quantities of a two- and three-dimensional Fermi gas from the operator-product expansion,''
\href{https://doi.org/10.1103/PhysRevA.84.043603}
{Phys.\ Rev.\ A \textbf{84}, 043603 (2011)}.

\bibitem{Goldberger:2012}
W.~D.~Goldberger and Z.~U.~Khandker,
``Viscosity sum rules at large scattering lengths,''
\href{https://doi.org/10.1103/PhysRevA.85.013624}
{Phys.\ Rev.\ A \textbf{85}, 013624 (2012)}.

\bibitem{Taylor:2012}
E.~Taylor and M.~Randeria,
``Apparent low-energy scale invariance in two-dimensional Fermi gases,''
\href{https://doi.org/10.1103/PhysRevLett.109.135301}
{Phys.\ Rev.\ Lett.\ \textbf{109}, 135301 (2012)}.

\bibitem{Massignan:2005}
P.~Massignan, G.~M.~Bruun, and H.~Smith,
``Viscous relaxation and collective oscillations in a trapped Fermi gas near the unitarity limit,''
\href{https://doi.org/10.1103/PhysRevA.71.033607}
{Phys.\ Rev.\ A \textbf{71}, 033607 (2005)}.

\bibitem{Bruun:2005}
G.~M.~Bruun and H.~Smith,
``Viscosity and thermal relaxation for a resonantly interacting Fermi gas,''
\href{https://doi.org/10.1103/PhysRevA.72.043605}
{Phys.\ Rev.\ A \textbf{72}, 043605 (2005)}.

\bibitem{Bruun:2012}
G.~M.~Bruun,
``Shear viscosity and spin-diffusion coefficient of a two-dimensional Fermi gas,''
\href{https://doi.org/10.1103/PhysRevA.85.013636}
{Phys.\ Rev.\ A \textbf{85}, 013636 (2012)}.

\bibitem{Schafer:2012}
T.~Sch\"afer,
``Shear viscosity and damping of collective modes in a two-dimensional Fermi gas,''
\href{https://doi.org/10.1103/PhysRevA.85.033623}
{Phys.\ Rev.\ A \textbf{85}, 033623 (2012)}.

\bibitem{Dusling:2013}
K.~Dusling and T.~Sch\"afer,
``Bulk viscosity and conformal symmetry breaking in the dilute Fermi gas near unitarity,''
\href{https://doi.org/10.1103/PhysRevLett.111.120603}
{Phys.\ Rev.\ Lett.\ \textbf{111}, 120603 (2013)}.

\bibitem{Chafin:2013}
C.~Chafin and T.~Sch\"afer,
``Scale breaking and fluid dynamics in a dilute two-dimensional Fermi gas,''
\href{https://doi.org/10.1103/PhysRevA.88.043636}
{Phys.\ Rev.\ A \textbf{88}, 043636 (2013)}.

\bibitem{Zwerger:2012}
W.~Zwerger (ed.),
\textit{The BCS-BEC Crossover and the Unitary Fermi Gas,}
\href{https://doi.org/10.1007/978-3-642-21978-8}
{Lecture Notes in Physics \textbf{836} (Springer, Berlin, Heidelberg, 2012)}.

\bibitem{Liu:2013}
X.-J.~Liu,
``Virial expansion for a strongly correlated Fermi system and its application to ultracold atomic Fermi gases,''
\href{https://doi.org/10.1016/j.physrep.2012.10.004}
{Phys.\ Rept.\ \textbf{524}, 37-83 (2013)}.

\bibitem{Fujii:2018}
K.~Fujii and Y.~Nishida,
``Hydrodynamics with spacetime-dependent scattering length,''
\href{https://doi.org/10.1103/PhysRevA.98.063634}
{Phys.\ Rev.\ A \textbf{98}, 063634 (2018)}.

\bibitem{Tan:2008a}
S.~Tan,
``Energetics of a strongly correlated Fermi gas,''
\href{https://doi.org/10.1016/j.aop.2008.03.004}
{Ann.\ Phys.\ (N.Y.) \textbf{323}, 2952-2970 (2008)}.

\bibitem{Tan:2008b}
S.~Tan,
``Large momentum part of a strongly correlated Fermi gas,''
\href{https://doi.org/10.1016/j.aop.2008.03.005}
{Ann.\ Phys.\ (N.Y.) \textbf{323}, 2971-2986 (2008)}.

\bibitem{Tan:2008c}
S.~Tan,
``Generalized virial theorem and pressure relation for a strongly correlated Fermi gas,''
\href{https://doi.org/10.1016/j.aop.2008.03.003}
{Ann.\ Phys.\ (N.Y.) \textbf{323}, 2987-2990 (2008)}.

\bibitem{Ho:2004}
T.-L.~Ho and E.~J.~Mueller,
``High temperature expansion applied to fermions near Feshbach resonance,''
\href{https://doi.org/10.1103/PhysRevLett.92.160404}
{Phys.\ Rev.\ Lett.\ \textbf{92}, 160404 (2004)}.

\bibitem{Yu:2009}
Z.~Yu, G.~M.~Bruun, and G.~Baym,
``Short-range correlations and entropy in ultracold-atom Fermi gases,''
\href{https://doi.org/10.1103/PhysRevA.80.023615}
{Phys.\ Rev.\ A \textbf{80}, 023615 (2009)}.

\bibitem{Braaten:2008}
E.~Braaten and L.~Platter,
``Exact relations for a strongly interacting Fermi gas from the operator product expansion,''
\href{https://doi.org/10.1103/PhysRevLett.100.205301}
{Phys.\ Rev.\ Lett.\ \textbf{100}, 205301 (2008)}.

\bibitem{Barth:2011}
M.~Barth and W.~Zwerger,
``Tan relations in one dimension,''
\href{https://doi.org/10.1016/j.aop.2011.05.010}
{Ann.\ Phys.\ \textbf{326}, 2544-2565 (2011)}.

\bibitem{Hofmann:2012}
J.~Hofmann,
``Quantum anomaly, universal relations, and breathing mode of a two-dimensional Fermi gas,''
\href{https://doi.org/10.1103/PhysRevLett.108.185303}
{Phys.\ Rev.\ Lett.\ \textbf{108}, 185303 (2012)}.

\bibitem{Martinez:2017}
M.~Martinez and T.~Sch\"afer,
``Hydrodynamic tails and a fluctuation bound on the bulk viscosity,''
\href{https://doi.org/10.1103/PhysRevA.96.063607}
{Phys.\ Rev.\ A \textbf{96}, 063607 (2017)}.

\bibitem{Gotze:1972}
W.~G\"otze and P. W\"olfle,
``Homogeneous dynamical conductivity of simple metals,''
\href{https://doi.org/10.1103/PhysRevB.6.1226}
{Phys.\ Rev.\ B \textbf{6}, 1226-1238 (1972)}.

\bibitem{Enss:2019}
T.~Enss,
``Dynamical viscosity, contact correlations and pairing in attractive Fermi gases,''
\href{https://arxiv.org/abs/1904.12772}
{arXiv:1904.12772 [cond-mat.quant-gas]}.

\bibitem{Hofmann:2019}
J.~Hofmann,
``High-temperature expansion of the viscosity in interacting quantum gases,''
\href{https://arxiv.org/abs/1905.05133}
{arXiv:1905.05133 [cond-mat.quant-gas]}.

\bibitem{Braaten:2006}
E.~Braaten and H.-W.~Hammer,
``Universality in few-body systems with large scattering length,''
\href{https://doi.org/10.1016/j.physrep.2006.03.001}
{Phys.\ Rep.\ \textbf{428}, 259-390 (2006)}.

\bibitem{Barth:2015}
M.~Barth and J.~Hofmann,
``Efimov correlations in strongly interacting Bose gases,''
\href{https://doi.org/10.1103/PhysRevA.92.062716}
{Phys.\ Rev.\ A \textbf{92}, 062716 (2015)}.

\end{thebibliography}
\end{document}